\documentclass[aps,prl,twocolumn,superscriptaddress,10pt, showpacs]{revtex4-2}

\usepackage{graphicx}
\usepackage{dcolumn}
\usepackage{amsmath}
\usepackage{nicefrac}
\usepackage{upgreek}
\usepackage{epstopdf}
\usepackage{color}
\usepackage[pagewise,modulo]{lineno}

\begin{document}

\title{Site- and Energy-Selective Low-Energy Electron Emission by X-Rays \\in Aqueous Phase}

\author{Dana Blo\ss}
\email{dana.bloss@uni-kassel.de}
	\affiliation{Institut f\"ur Physik und CINSaT, Universit\"at Kassel, Heinrich-Plett-Straße\,40, 34132 Kassel, Germany}

\author{R\'{e}mi Dupuy}
	\affiliation{Laboratoire de Chimie Physique - Mati\`{e}re et Rayonnement, Sorbonne Universit\'{e}, CNRS, LCP-MR, 75005 Paris Cedex 05, France}
	
\author{Florian Trinter}
	\affiliation{Fritz-Haber-Institut der Max-Planck-Gesellschaft, Faradayweg 4-6, 14195 Berlin, Germany}
	\affiliation{Institut f\"ur Kernphysik, Goethe-Universit\"at Frankfurt, Max-von-Laue-Straße\,1, 60438 Frankfurt am Main, Germany}
	
\author{Isaak Unger}
	\affiliation{Department of Physics and Astronomy, Uppsala University, Box 516, 75120 Uppsala, Sweden}
	
\author{Noelle Walsh}
	\affiliation{MAX IV Laboratory, Lund University, Box 118, 22100 Lund, Sweden}
	
\author{\\Gunnar \"{O}hrwall}
	\affiliation{MAX IV Laboratory, Lund University, Box 118, 22100 Lund, Sweden}

\author{Niklas Golchert}
	\affiliation{Institut f\"ur Physik und CINSaT, Universit\"at Kassel, Heinrich-Plett-Straße\,40, 34132 Kassel, Germany}

\author{Gabriel Klassen}
	\affiliation{Institut f\"ur Physik und CINSaT, Universit\"at Kassel, Heinrich-Plett-Straße\,40, 34132 Kassel, Germany}
\author{Adrian Krone}
	\affiliation{Institut f\"ur Physik und CINSaT, Universit\"at Kassel, Heinrich-Plett-Straße\,40, 34132 Kassel, Germany}

\author{Yusaku Terao}
		\affiliation{Institut f\"ur Physik und CINSaT, Universit\"at Kassel, Heinrich-Plett-Straße\,40, 34132 Kassel, Germany}
		
\author{\\Johannes H. Viehmann}
	\affiliation{Institut f\"ur Physik und CINSaT, Universit\"at Kassel, Heinrich-Plett-Straße\,40, 34132 Kassel, Germany}
	
\author{Lasse W\"{u}lfing}
	\affiliation{Fakult\"{a}t Physik, Technische Universit\"{a}t Dortmund, Maria-Goeppert-Mayer-Str. 2, 44227 Dortmund, Germany}

\author{Clemens Richter}
	\affiliation{Fritz-Haber-Institut der Max-Planck-Gesellschaft, Faradayweg 4-6, 14195 Berlin, Germany}

\author{Tillmann Buttersack}
	\affiliation{Fritz-Haber-Institut der Max-Planck-Gesellschaft, Faradayweg 4-6, 14195 Berlin, Germany}

\author{Lorenz S. Cederbaum}
	\affiliation{Theoretische Chemie, Institut f\"ur Physikalische Chemie, Universit\"at Heidelberg, Im Neuenheimer Feld 229, 69120 Heidelberg, Germany}
	
\author{\\Uwe Hergenhahn}
	\affiliation{Fritz-Haber-Institut der Max-Planck-Gesellschaft, Faradayweg 4-6, 14195 Berlin, Germany}

\author{Olle Bj\"orneholm}
	\affiliation{Department of Physics and Astronomy, Uppsala University, Box 516, 75120 Uppsala, Sweden}
	
\author{Arno Ehresmann}
	\affiliation{Institut f\"ur Physik und CINSaT, Universit\"at Kassel, Heinrich-Plett-Straße\,40, 34132 Kassel, Germany}

\author{Andreas Hans}
\email{hans@physik.uni-kassel.de}
	\affiliation{Institut f\"ur Physik und CINSaT, Universit\"at Kassel, Heinrich-Plett-Straße\,40, 34132 Kassel, Germany}

\date{\today}

\begin{abstract}
Low-energy-electron emission from resonant Auger final states via intermolecular Coulombic decay (RA-ICD) has been previously described as a promising scenario for controlling radiation damage for medical purposes, but has so far only been observed in prototypical atomic and molecular van der Waals dimers and clusters. Here, we report the experimental observation of RA-ICD in aqueous solution. We show that for solvated Ca\textsuperscript{2+} ions, the emission can be very efficiently controlled by tuning the photon energy of exciting X-rays to inner-shell resonances of the ions. Our results provide the next step from proving RA-ICD in relatively simple prototype systems to understanding the relevance and potential applications of ICD in real-life scenarios.

\end{abstract}

\maketitle

A major challenge in x-ray-based radiation therapies is to achieve the largest possible contrast between the doses of radiation deposited in the malignant versus the surrounding healthy tissue, which the radiation inevitably needs to pass through. About a decade ago, it was suggested that this contrast could be significantly enhanced by making use of a novel mechanism which locally produces radicals and low-energy electrons (LEEs), namely intermolecular Coulombic decay (ICD) of states populated by resonant Auger (RA) decay \cite{Gokhberg.2014, Trinter.2014}. The reaction products of the RA-ICD process are presumably the main mediators of biological damage by causing single or multiple DNA strand breaks \cite{Boudaiffa.2000,Alizadeh.2012, Alizadeh.2015}.

ICD  is a non-local autoionization mechanism in which the excess energy of an exited part of an extended system is transferred to a neighbor, thereby ionizing it \cite{Cederbaum.1997, Marburger.2003, Jahnke.2004, Jahnke.2020}. First predicted in 1997 \cite{Cederbaum.1997}, ICD and a number of related processes have by now been observed in a plethora of systems \cite{Jahnke.2020}. The initial excitation of the system can proceed via different mechanisms. The vast majority of experimental reports, however, uses photoionization \cite{Jahnke.2020}. 

\begin{figure}[h!]
\includegraphics[width=.48\textwidth]{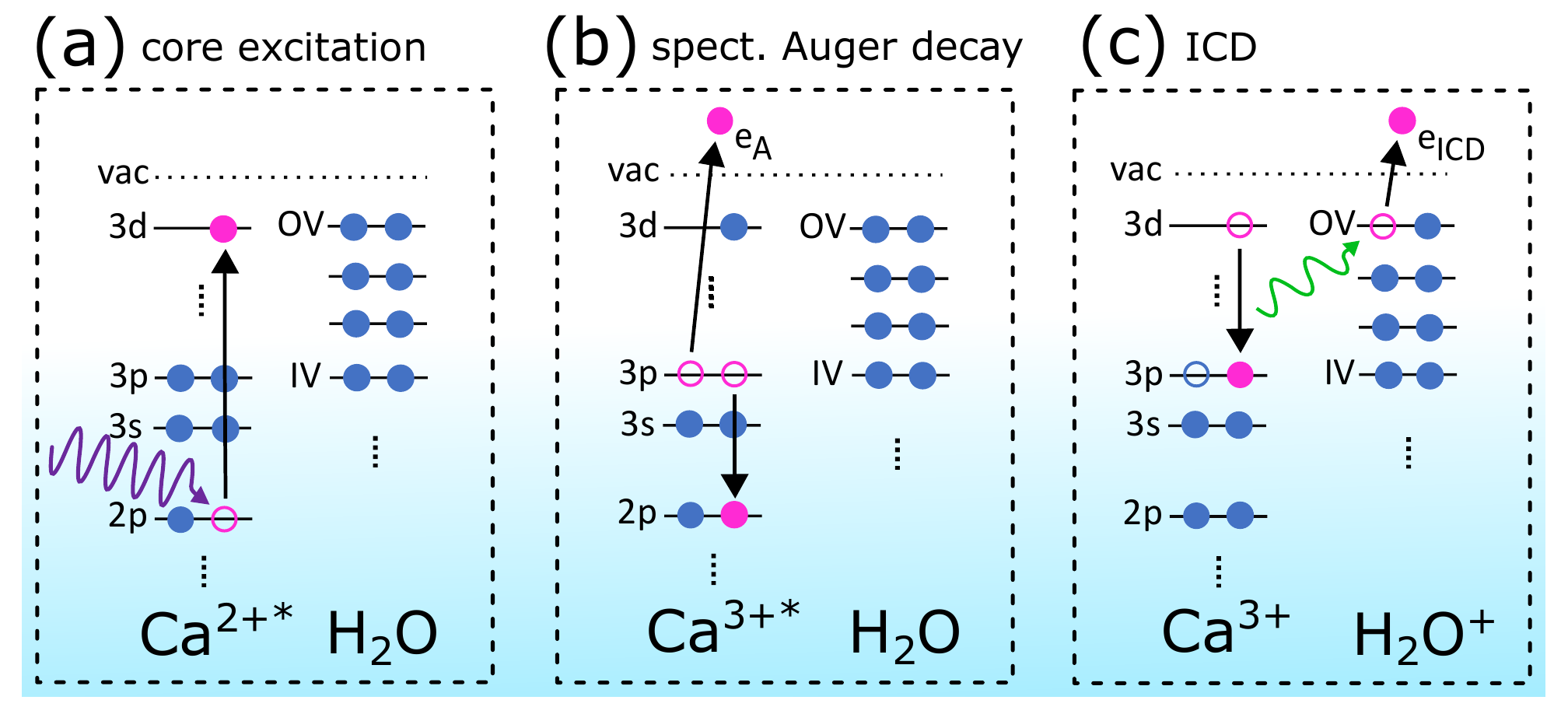}
  \caption{Illustration of the process of interest. (a) Resonant photoexcitation of a Ca$^{2+}$ ion. (b) Spectator resonant Auger decay. (c) Interatomic Coulombic decay with ionization of a neighboring water molecule. For simplicity, only one water neighbor is shown, but in the experiments the Ca$^{2+}$ ion is fully solvated. Dashed vertical lines indicate energetically lower lying electronic levels, which are not relevant for the present discussion.}
  \label{fig:sketch}
\end{figure}

The particular variant of interest, RA-ICD, is illustrated in Fig.~\ref{fig:sketch}. In this scenario, an atom or molecule, which is weakly bound to one or more neighbors, e.g., through van der Waals or hydrogen bonds, is first resonantly inner-shell-excited by X-rays [Fig.~\ref{fig:sketch}(a)]. The resonant nature of this first step of the overall mechanism ensures a largely higher absorption cross section as compared to absorption due to photoionization into a continuum. A common decay path of such core-excited species is spectator RA decay. Here, a valence electron fills the inner-shell vacancy and another valence electron is emitted, while the initially excited electron stays as a ``spectator'' in its orbital [Fig.~\ref{fig:sketch}(b)]. The internal excess energy of the RA final states is typically not enough for further local autoionization. If the atom or molecule does not have neighbors, further decay is only possible radiatively and/or via dissociation. The presence of neighbors, however, enables further autoionization of the system as a whole via ICD: the excited ion decays to its ground state transferring the excess energy to ionize a valence electron from a neighbor [Fig.~\ref{fig:sketch}(c)]. The emitted electron, the so-called ICD electron, has typically a rather low kinetic energy below 30~eV. A potential relevance of ICD and related phenomena in radiation biology has been recently discussed intensely \cite{Mucke.2010, Hergenhahn.2012, Jahnke.2020}.

From the viewpoint of medical applications, RA-ICD provides an intriguing advantage. Resonant inner-shell excitation with X-rays is highly element-selective, meaning that at a particular resonance energy, photons are nearly exclusively absorbed by one species of atom, while the rest of a system is basically transparent. This can be utilized for targeted energy deposition, e.g., via a marker element, while leaving the surroundings mostly unaffected. The destructive LEEs and radical are thus produced locally at the point of interest. In Fig.~\ref{fig:sketch}, the process is exemplified for a Ca$^{2+}$ ion with a water molecule as its neighbor, the case which will be studied in the present work.

Soon after RA-ICD and its potential for medical applications was recognized in a theoretical work \cite{Gokhberg.2014} and in an experimental study on molecular van der Waals dimers \cite{Trinter.2014}, it has also been observed in Ar dimers \cite{Kimura.2013}. Simultaneously, it was shown that in heterogeneous rare-gas dimers the energies of the emitted LEEs can be adjusted by choosing a neighbor atom with appropriate ionization energy \cite{Kimura.2013b, Miteva.2014, OKeeffe2014}. While ICD from non-resonant Auger final states has been the topic of various other studies before and after that \cite{Cederbaum.1997, Jahnke.2020}, no attempt has been reported to transfer RA-ICD from prototypical van der Waals dimers to more realistic samples. A major challenge for the investigation of ICD and related mechanisms in larger and more complex systems, e.g., in liquids, is the overwhelming background of LEEs due to other processes, mainly electron-impact excitation or ionization and quasi-elastic scattering processes \cite{Malerz.2021}. In the present study, we overcome this challenge by applying electron-electron coincidence spectroscopy to an aqueous solution and report the first observation of RA-ICD in liquids.

As an explicit example, we consider the $2p\rightarrow 3d$ resonant excitation of solvated Ca$^{2+}$ ions. To obtain information for both the high-resolution resonant Auger spectra as well as the subsequently emitted LEEs, we combine the results of two independently performed experiments (see Methods section). 

\begin{figure}[h!]
\includegraphics[width=.45\textwidth]{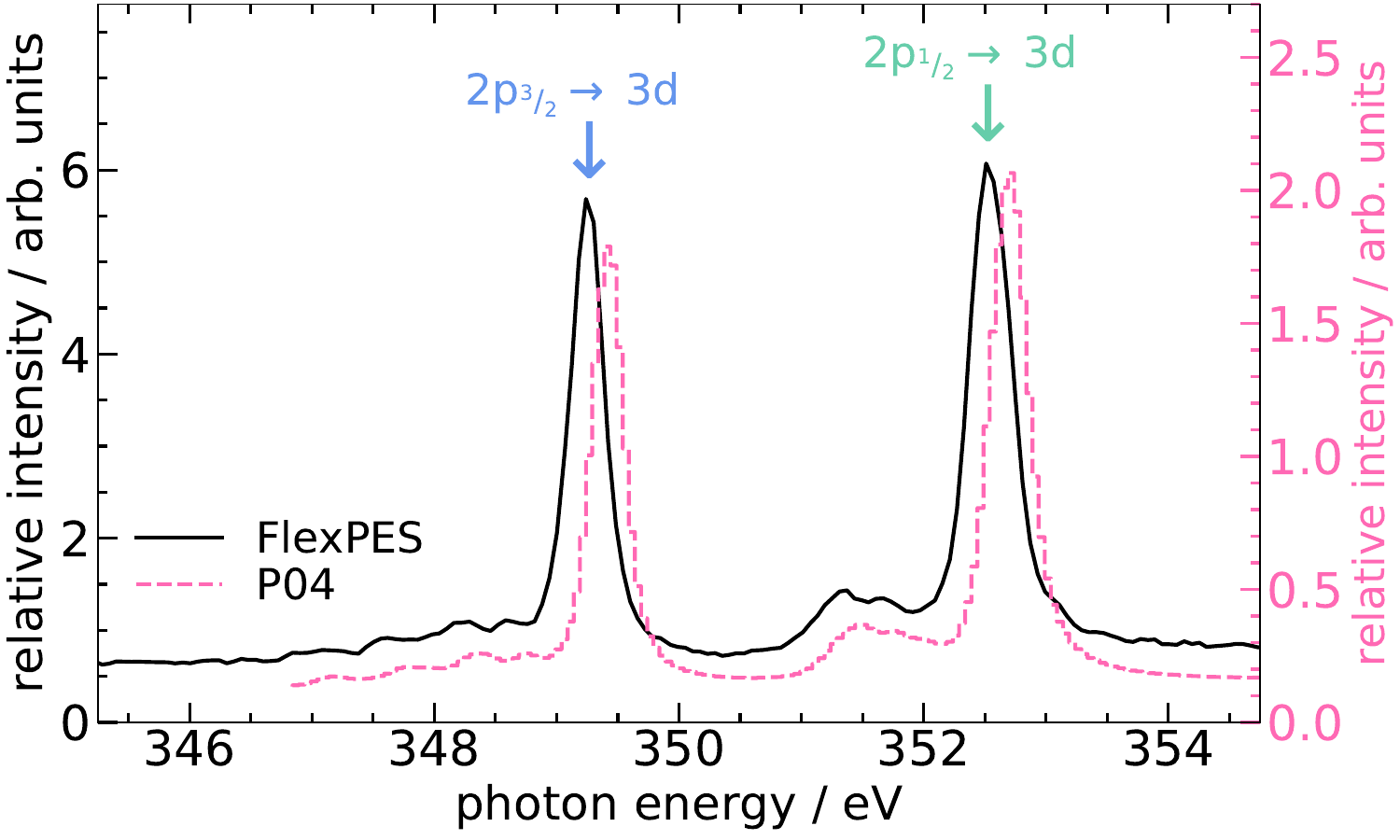}
  \caption{Total electron yield (FlexPES, black solid line and left $y$ axis) and partial electron yield (P04, magenta dashed line and right $y$ axis) as a function of the exciting-photon energy across the Ca$^{2+}~2p$ edge from both conducted experiments. A slight shift is observed between the experiments at P04 and FlexPES, respectively, which is most likely a result of energy-calibration uncertainties (see text).}
  \label{fig:eyield}
\end{figure}

In both experiments, the energetic positions of the Ca$^{2+}$ $2p\rightarrow 3d$ resonances were identified by scanning the exciting-photon energy stepwise and recording the total or partial electron yield. Typical electron yield curves are shown in Fig.~\ref{fig:eyield}. Two prominent resonances attributed to the $2p_{\nicefrac{1}{2},\nicefrac{3}{2}}\rightarrow 3d$ fine-structure doublet are present~\cite{Abid.2021}. The weaker side structures, assigned to crystal field effects, have been discussed elsewhere \cite{Abid.2021, Yang.2020, Rubensson.1994}. Note that even after careful photon-energy calibration, a slight deviation on the order of 200~meV in resonance energies remains between the two experiments, with neither exactly matching previously reported values \cite{Abid.2021}. The origin of this deviation cannot be reconstructed. It is, however, most likely a result of the fact that the calibrations were done not exactly in the energy range of the Ca L-edge and can be regarded as uncertainty for the observed resonance energies. Since the resonances were identified in each experiment individually, this does not affect our conclusions in any way. 

\begin{figure}[h!]
\includegraphics[width=.48\textwidth]{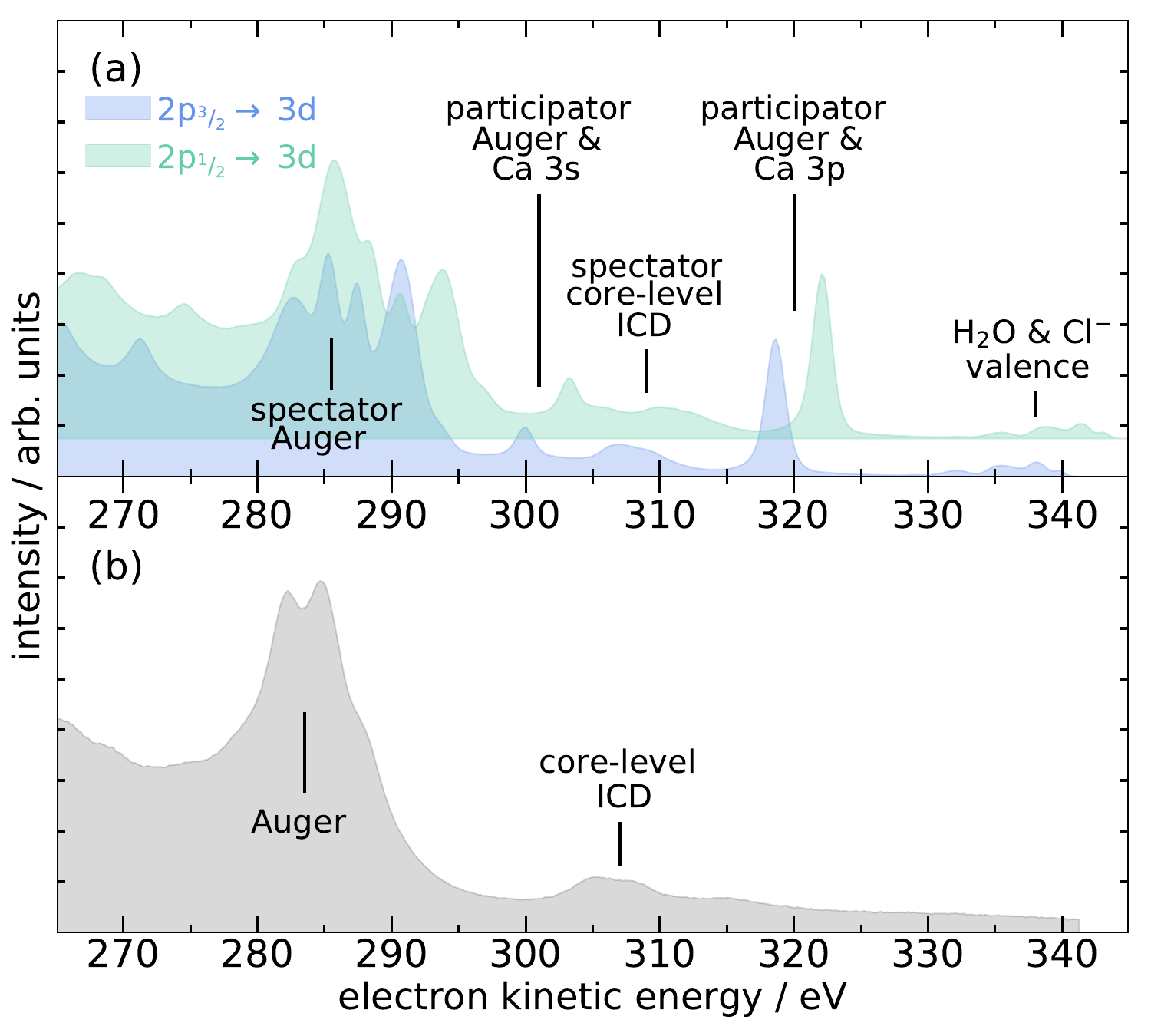}
  \caption{(a) Resonant Auger spectra after Ca$^{2+}$ $2p_{\nicefrac{3}{2}}\rightarrow 3d$ (349.4~eV, blue) and $2p_{\nicefrac{1}{2}}\rightarrow 3d$ (352.7~eV, green) photoexcitation. An offset has been applied to the latter for better visibility. (b) Auger spectrum after $2p$ photoionization, recorded at 460~eV exciting-photon energy. The main features are labeled, for detailed discussion see text.}
  \label{fig:Augerspectra}
\end{figure}

Using high-resolution electron spectroscopy, the resonant Auger spectra on both resonances and well above the $2p$ edge (binding energies of 352.8~eV and 356.6~eV for the  $2p_{\nicefrac{3}{2}}$ and $2p_{\nicefrac{1}{2}}$ components \cite{Pokapanich.2011}) were measured and are displayed in Fig.~\ref{fig:Augerspectra}. The non-resonant Auger spectrum, shown in Fig.~\ref{fig:Augerspectra}(b), has been reported and discussed earlier \cite{Pokapanich.2011,Ottosson.2012} and serves as a reference for the interpretation of the spectra recorded on the resonances. It contains a main contribution of conventional Auger electrons between 280~eV and 290~eV and a weaker signal between 300~eV and 310~eV attributed to core-level ICD. This latter variant of ICD is a direct competitor to Auger decay, a valence electron from Ca fills the core vacancy and a valence electron from a neighboring water molecule is emitted \cite{Pokapanich.2011}. Relative to the non-resonant Auger spectrum, the features in the two spectra on the resonances are rather straightforward to assign. The fastest electrons in the range from 330~eV to about 340~eV are mainly valence electrons from water. The sharp prominent peaks at 318.7~eV ($2p_{\nicefrac{3}{2}}$ resonance) and 322.1~eV ($2p_{\nicefrac{1}{2}}$ resonance) represent the  $3p$ photoelectrons from Ca$^{2+}$ with a binding energy of 29.8~eV~\cite{Pokapanich.2011}. Their intensity is resonantly enhanced due to participator resonant Auger decay. At about 19~eV lower kinetic energy, another relatively sharp peak can be attributed to the Ca $3s$ photoelectrons. In between the $3s$ and $3p$ photoelectrons, there is a broader peak which originates from resonant core-level ICD, which has recently been analyzed in detail \cite{Dupuy.2024}. In the region between about 282~eV and 295~eV, the spectator resonant Auger spectrum corresponding to transitions from Ca$^{2+}$($2p^{-1}3d$) to Ca$^{3+}$($3p^{-2}3d$) configurations can be observed, with a characteristic ``spectator shift'' \cite{Ottosson.2012} compared to the non-resonant Auger spectrum. At even lower kinetic energies transitions to final states with holes in the $3s$ level occur.

The remaining internal energy in the Ca$^{3+}$ ion can easily be estimated from the difference between the spectator resonant Auger region and the $3p$ photoelectron peak in the kinetic-energy spectrum, which represents the ionic ground state. This energy amounts to about 28 to 37~eV and is available to be transferred to neighbors. The kinetic energies of electrons emitted in ICD of these states can then be coarsely calculated. The ionization energy of water (11.3~eV in liquid phase \cite{Winter.2023}) and the potential Coulomb energy between the resulting ions need to be subtracted from the available excess energy. The Coulomb energy between the Ca$^{3+}$ and the H$_2$O$^+$ ion strongly depends on their inter-ionic distance and the screening by the environment. From the average Ca-O distance of 2.46~\AA\ \cite{Jalilehvand.2001} a nominal Coulomb energy of 17.6~eV can be calculated \cite{Pokapanich.2011}, while at infinite distance or for complete screening it is zero. Experimentally, the Coulomb energy can be deduced from the energetic distance between Auger and core-level ICD features in previous studies or from the present kinetic energies in the spectra in Fig.~\ref{fig:Augerspectra}. It turns out that there is almost perfect screening by the environment and that the two-site ionization potentials of the ICD final states are close to the sum of the ionization energies of the individual ions \cite{Pokapanich.2011}. Consequently, a Coulomb energy of at most a few eV needs to be considered. Note that the final ionic states of core-level ICD and RA-ICD are identical, namely Ca$^{3+}$ and H$_2$O$^+$. Under these considerations we expect an RA-ICD spectrum centered at about 20~eV and of about 10~eV width. Except for slight changes in the spectral structure the energy range is expected to be independent from the fine-structure component, since this difference is mainly taken by the spectator Auger electron. 

\begin{figure}[h!]
\includegraphics[width=.48\textwidth]{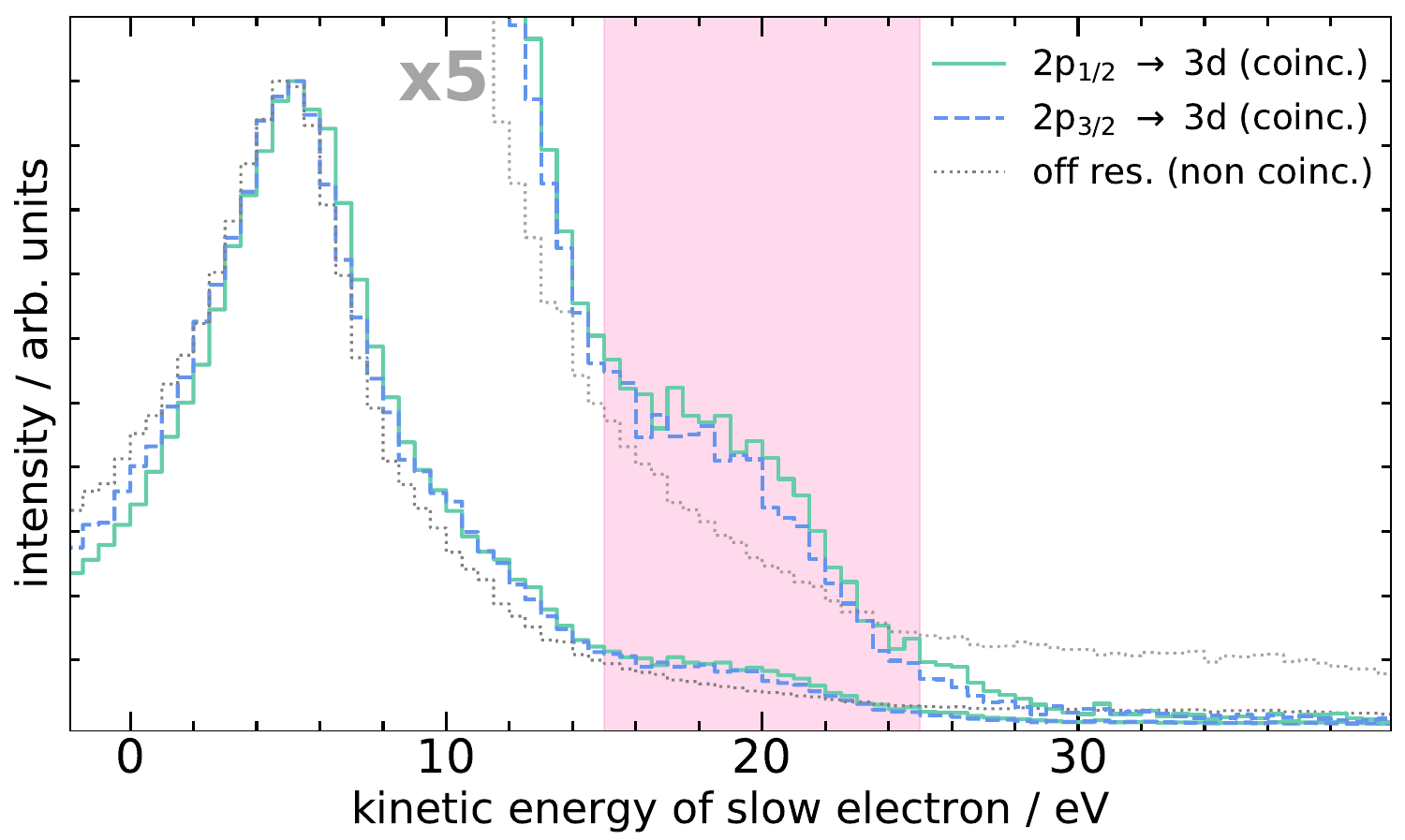}
  \caption{LEE spectra of a 4~M CaCl$_2$ solution exposed to X-rays of different energies. The gray dotted curve is a reference spectrum of single-electron events above the Ca$^{2+}$~$2p$ ionization threshold taken at 405~eV exciting-photon energy. The blue dashed line and the green solid line represent spectra on the $2p_{\nicefrac{3}{2}}\rightarrow 3d$ and $2p_{\nicefrac{1}{2}}\rightarrow 3d$ resonances at 349.25~eV and 352.5~eV, respectively. The spectra on the resonances were filtered using the coincidence condition that the first of two coincidentally detected electrons is the Ca$^{2+}$~$2p\rightarrow 3d$ spectator resonant Auger electron. All spectra are normalized to their maximum and are additionally plotted magnified to emphasize the low-intensity region. The red shaded area indicates the expected range of the RA-ICD electrons.}
 \label{fig:ICD}
\end{figure}

This estimated kinetic-energy range of RA-ICD electrons lies energetically within the typical background of LEEs observed in electron spectra from liquids \cite{Malerz.2021}. Without further discrimination, the identification of a broad feature in this range is extremely challenging using conventional electron spectroscopy. We therefore applied electron-electron coincidence spectroscopy to reduce the background of inelastically scattered electrons tremendously. Due to the large photoexcitation cross section on the Ca$^{2+}$ $2p$ resonances, the resonant Auger electrons in the range between 282 and 295~eV can be unambiguously identified in the electron spectrum \cite{SM}, although the resolution is significantly inferior to the hemispherical analyzer. We then select events of two-electron coincidences (resonant Auger electron and RA-ICD electron). 
In Fig.~\ref{fig:ICD}, the spectra of the second, slower electron are plotted for both resonances with the coincidence condition that the first electron was the spectator Auger electron. The off-resonant spectrum of single-electron events serves as a reference. The spectra have been normalized to their maximum to allow for comparison of spectral features within a single graph and additionally are plotted in a magnified presentation to emphasize features in the region of lower intensity above 10~eV. The expected energy range (15-25~eV) of the RA-ICD electrons has been highlighted. For a more detailed description of the coincidence data treatment see the Supplemental Material \cite{SM}.

It is evident that also the coincidence filtering cannot completely suppress the LEE tail increasing towards zero kinetic energy, which is omnipresent in the spectrum of the second electron. Note that the spectral shape close to zero~eV may not be accurate due to the difficulty of measuring very slow electrons from a liquid \cite{Malerz.2021}. Importantly, however, on top of the LEE background, for both resonances a significant structure can be observed which perfectly matches the estimated RA-ICD kinetic-energy range and is absent in the reference spectrum. We thus conclude that the observed feature between about 15 and 25~eV kinetic energy originates from RA-ICD of solvated Ca$^{3+}$ ions. A second feature may be identified around 10~eV, which gradually merges into the LEE background and is therefore less discernible. Its origin cannot be unambiguously identified from the present data and could possibly include RA-ICD ionizing water valence orbitals with higher binding energies, or electron-transfer-mediated decay of Ca$^{3+}$ ions subsequent to RA-ICD \cite{Blo.2024, Gopakumar.2023}.

For van der Waals-bound dimers, RA-ICD was found to be very efficient, occurring on a timescale below 20~fs~\cite{Trinter.2014}. A similar time range was reported for direct core-level ICD of $2p$-ionized solvated Ca ions \cite{Pokapanich.2011}. ICD of valence-ionized states in Mg ions was predicted to be even faster, below 1~fs \cite{Stumpf.2016}. Although none of these cases is equivalent to the presently discussed variant of ICD and a quantification is not possible from the present data, it is reasonable to assume a similar lifetime here.

We emphasize the resonant and site-selective character of the LEE emission triggered by RA-ICD, which was the fundamental idea in Refs.~\cite{Gokhberg.2014, Trinter.2014} to suggest its application for medical purposes. Through the resonant excitation as a precondition for RA-ICD, the production of LEEs and radicals should only appear in a very narrow window of the x-ray photon energy (sub 1~eV wide in the present case). Naturally, the LEEs originating from RA-ICD are only emitted locally from the immediate surroundings of the Ca ions. While Ca is abundant in the biosphere, the scenario could be transferred to another marker element, which may artificially be introduced into an organism, enriched, e.g., in cancer tissue. Choosing the resonant photon energy of the marker element, destructive LEEs can then be produced efficiently and locally, while the rest of the tissue is basically transparent for the X-rays. Higher atomic number $Z$ will increase the contrast/selectivity, as the photoionization cross section for C, N, and O drops with increasing photon energy. Moreover, deep core-level vacancies in high-Z elements give rise to more electrons due to cascade decay \cite{Blo.2024}. It should also be noted that the damage caused by LEEs depends on their kinetic energy distribution. For example, the impact of electrons with different kinetic energies may mainly cause single, double, or multiple strand breaks to DNA \cite{Boudaiffa.2000, Orlando.2008, Alizadeh.2012, Alizadeh.2015}. In general, the distribution of kinetic energies of electrons emitted due to RA-ICD can be steered by choosing an appropriate system.

\section{Methods}
High-resolution Auger spectra were recorded using a hemispherical electron analyzer coupled to a liquid microjet device \cite{Winter.2006, Malerz.2022}. The instrument was installed at the P04 beamline \cite{Viefhaus.2013} of the PETRA~III synchrotron facility (DESY, Hamburg, Germany). A 1.5~M solution of CaCl$_2$ was prepared by dissolving a commercial salt (Sigma-Aldrich, 99\% purity) in ultra-pure water (resistivity 18.2~M$\Upomega$) and subsequent filtration and degassing. For typical operation conditions of the liquid microjet we used 30~$\upmu$m diameter glass nozzles and a flow rate of 0.8~mL/min. All spectra were recorded by applying a $-50$~V bias voltage to the liquid jet, allowing to suppress water gas-phase contributions to the spectra \cite{Winter.2023}. The photon energy of the beamline was calibrated against known gas-phase K-edge absorption lines of different gases (CO, N$_2$, SF$_6$, and Ne). We operated with a photon bandwidth of about 60~meV and an analyzer resolution of 200~meV, which is sufficient to resolve all features in the Auger spectra.

For monitoring LEE emission occurring subsequently to resonant Auger decay we used a setup for electron-electron coincidence spectroscopy \cite{Pohl.2017, Blo.2024}. This experiment was performed at the FlexPES beamline at the MAX~IV laboratory in Lund, Sweden \cite{Preobrajenski.2023} during single-bunch delivery. A magnetic-bottle time-of-flight electron spectrometer was used to efficiently record coincidences of two electrons emitted in pairs after a single excitation. Details of the setup are described elsewhere \cite{Pohl.2017, Blo.2024}. A 4~M solution of CaCl$_2$ was prepared by the same procedure as in the first experiment. The sample was introduced into vacuum at a temperature of 4~°C and at flow rates between 0.6-0.8~mL/min. For measuring the LEE spectra, a $+26$~V bias voltage was applied to the drift tube of the magnetic bottle in order to accelerate electrons with near zero kinetic energy such that they arrive at the detector within the time window between two consecutive pulses (320~ns) and a $-3$~V bias voltage was applied to the jet itself. The exciting-photon energy axis was calibrated at the Ar L-edge and the time-of-flight axis was converted to kinetic energies by measuring water O~1s photoemission (for which the binding energy is known) at various exciting-photon energies. Using a procedure of data acquisition during several consecutive exciting-photon pulses, the relevance of random coincidences was estimated to be negligible \cite{SM}.

\subsection{Data availability}
The data generated in this study have been deposited in a Zenodo database [https://doi.org/10.0.20.161/zenodo.13358738].

\subsection{Code availability}
For the data evaluation freely available, common python packages were used. The developed code is available from the corresponding authors upon request.

\bibliography{manuscript}

\begin{thebibliography}{33}%
\makeatletter
\providecommand \@ifxundefined [1]{%
 \@ifx{#1\undefined}
}%
\providecommand \@ifnum [1]{%
 \ifnum #1\expandafter \@firstoftwo
 \else \expandafter \@secondoftwo
 \fi
}%
\providecommand \@ifx [1]{%
 \ifx #1\expandafter \@firstoftwo
 \else \expandafter \@secondoftwo
 \fi
}%
\providecommand \natexlab [1]{#1}%
\providecommand \enquote  [1]{``#1''}%
\providecommand \bibnamefont  [1]{#1}%
\providecommand \bibfnamefont [1]{#1}%
\providecommand \citenamefont [1]{#1}%
\providecommand \href@noop [0]{\@secondoftwo}%
\providecommand \href [0]{\begingroup \@sanitize@url \@href}%
\providecommand \@href[1]{\@@startlink{#1}\@@href}%
\providecommand \@@href[1]{\endgroup#1\@@endlink}%
\providecommand \@sanitize@url [0]{\catcode `\\12\catcode `\$12\catcode
  `\&12\catcode `\#12\catcode `\^12\catcode `\_12\catcode `\%12\relax}%
\providecommand \@@startlink[1]{}%
\providecommand \@@endlink[0]{}%
\providecommand \url  [0]{\begingroup\@sanitize@url \@url }%
\providecommand \@url [1]{\endgroup\@href {#1}{\urlprefix }}%
\providecommand \urlprefix  [0]{URL }%
\providecommand \Eprint [0]{\href }%
\providecommand \doibase [0]{https://doi.org/}%
\providecommand \selectlanguage [0]{\@gobble}%
\providecommand \bibinfo  [0]{\@secondoftwo}%
\providecommand \bibfield  [0]{\@secondoftwo}%
\providecommand \translation [1]{[#1]}%
\providecommand \BibitemOpen [0]{}%
\providecommand \bibitemStop [0]{}%
\providecommand \bibitemNoStop [0]{.\EOS\space}%
\providecommand \EOS [0]{\spacefactor3000\relax}%
\providecommand \BibitemShut  [1]{\csname bibitem#1\endcsname}%
\let\auto@bib@innerbib\@empty
\bibitem [{\citenamefont {Gokhberg}\ \emph {et~al.}(2014)\citenamefont
  {Gokhberg}, \citenamefont {Koloren{\v{c}}}, \citenamefont {Kuleff},\ and\
  \citenamefont {Cederbaum}}]{Gokhberg.2014}%
  \BibitemOpen
  \bibfield  {author} {\bibinfo {author} {\bibfnamefont {K.}~\bibnamefont
  {Gokhberg}}, \bibinfo {author} {\bibfnamefont {P.}~\bibnamefont
  {Koloren{\v{c}}}}, \bibinfo {author} {\bibfnamefont {A.~I.}\ \bibnamefont
  {Kuleff}},\ and\ \bibinfo {author} {\bibfnamefont {L.~S.}\ \bibnamefont
  {Cederbaum}},\ }\href {https://doi.org/10.1038/nature12936} {\bibfield
  {journal} {\bibinfo  {journal} {Nature}\ }\textbf {\bibinfo {volume} {505}},\
  \bibinfo {pages} {661} (\bibinfo {year} {2014})}\BibitemShut {NoStop}%
\bibitem [{\citenamefont {Trinter}\ \emph {et~al.}(2014)\citenamefont
  {Trinter}, \citenamefont {Sch{\"o}ffler}, \citenamefont {Kim}, \citenamefont
  {Sturm}, \citenamefont {Cole}, \citenamefont {Neumann}, \citenamefont
  {Vredenborg}, \citenamefont {Williams}, \citenamefont {Bocharova},
  \citenamefont {Guillemin}, \citenamefont {Simon}, \citenamefont {Belkacem},
  \citenamefont {Landers}, \citenamefont {Weber}, \citenamefont
  {Schmidt-B{\"o}cking}, \citenamefont {D{\"o}rner},\ and\ \citenamefont
  {Jahnke}}]{Trinter.2014}%
  \BibitemOpen
  \bibfield  {author} {\bibinfo {author} {\bibfnamefont {F.}~\bibnamefont
  {Trinter}}, \bibinfo {author} {\bibfnamefont {M.~S.}\ \bibnamefont
  {Sch{\"o}ffler}}, \bibinfo {author} {\bibfnamefont {H.-K.}\ \bibnamefont
  {Kim}}, \bibinfo {author} {\bibfnamefont {F.~P.}\ \bibnamefont {Sturm}},
  \bibinfo {author} {\bibfnamefont {K.}~\bibnamefont {Cole}}, \bibinfo {author}
  {\bibfnamefont {N.}~\bibnamefont {Neumann}}, \bibinfo {author} {\bibfnamefont
  {A.}~\bibnamefont {Vredenborg}}, \bibinfo {author} {\bibfnamefont
  {J.}~\bibnamefont {Williams}}, \bibinfo {author} {\bibfnamefont
  {I.}~\bibnamefont {Bocharova}}, \bibinfo {author} {\bibfnamefont
  {R.}~\bibnamefont {Guillemin}}, \bibinfo {author} {\bibfnamefont
  {M.}~\bibnamefont {Simon}}, \bibinfo {author} {\bibfnamefont
  {A.}~\bibnamefont {Belkacem}}, \bibinfo {author} {\bibfnamefont {A.~L.}\
  \bibnamefont {Landers}}, \bibinfo {author} {\bibfnamefont {T.}~\bibnamefont
  {Weber}}, \bibinfo {author} {\bibfnamefont {H.}~\bibnamefont
  {Schmidt-B{\"o}cking}}, \bibinfo {author} {\bibfnamefont {R.}~\bibnamefont
  {D{\"o}rner}},\ and\ \bibinfo {author} {\bibfnamefont {T.}~\bibnamefont
  {Jahnke}},\ }\href {https://doi.org/10.1038/nature12927} {\bibfield
  {journal} {\bibinfo  {journal} {Nature}\ }\textbf {\bibinfo {volume} {505}},\
  \bibinfo {pages} {664} (\bibinfo {year} {2014})}\BibitemShut {NoStop}%
\bibitem [{\citenamefont {Bouda{\"i}ffa}\ \emph {et~al.}(2000)\citenamefont
  {Bouda{\"i}ffa}, \citenamefont {Cloutier}, \citenamefont {Hunting},
  \citenamefont {Huels},\ and\ \citenamefont {Sanche}}]{Boudaiffa.2000}%
  \BibitemOpen
  \bibfield  {author} {\bibinfo {author} {\bibfnamefont {B.}~\bibnamefont
  {Bouda{\"i}ffa}}, \bibinfo {author} {\bibfnamefont {P.}~\bibnamefont
  {Cloutier}}, \bibinfo {author} {\bibfnamefont {D.}~\bibnamefont {Hunting}},
  \bibinfo {author} {\bibfnamefont {M.~A.}\ \bibnamefont {Huels}},\ and\
  \bibinfo {author} {\bibfnamefont {L.}~\bibnamefont {Sanche}},\ }\href
  {https://doi.org/10.1126/science.287.5458.1658} {\bibfield  {journal}
  {\bibinfo  {journal} {Science}\ }\textbf {\bibinfo {volume} {287}},\ \bibinfo
  {pages} {1658} (\bibinfo {year} {2000})}\BibitemShut {NoStop}%
\bibitem [{\citenamefont {Alizadeh}\ and\ \citenamefont
  {Sanche}(2012)}]{Alizadeh.2012}%
  \BibitemOpen
  \bibfield  {author} {\bibinfo {author} {\bibfnamefont {E.}~\bibnamefont
  {Alizadeh}}\ and\ \bibinfo {author} {\bibfnamefont {L.}~\bibnamefont
  {Sanche}},\ }\href {https://doi.org/10.1021/cr300063r} {\bibfield  {journal}
  {\bibinfo  {journal} {Chemical Reviews}\ }\textbf {\bibinfo {volume} {112}},\
  \bibinfo {pages} {5578} (\bibinfo {year} {2012})}\BibitemShut {NoStop}%
\bibitem [{\citenamefont {Alizadeh}\ \emph {et~al.}(2015)\citenamefont
  {Alizadeh}, \citenamefont {Orlando},\ and\ \citenamefont
  {Sanche}}]{Alizadeh.2015}%
  \BibitemOpen
  \bibfield  {author} {\bibinfo {author} {\bibfnamefont {E.}~\bibnamefont
  {Alizadeh}}, \bibinfo {author} {\bibfnamefont {T.~M.}\ \bibnamefont
  {Orlando}},\ and\ \bibinfo {author} {\bibfnamefont {L.}~\bibnamefont
  {Sanche}},\ }\href {https://doi.org/10.1146/annurev-physchem-040513-103605}
  {\bibfield  {journal} {\bibinfo  {journal} {Annual Review of Physical
  Chemistry}\ }\textbf {\bibinfo {volume} {66}},\ \bibinfo {pages} {379}
  (\bibinfo {year} {2015})}\BibitemShut {NoStop}%
\bibitem [{\citenamefont {Cederbaum}\ \emph {et~al.}(1997)\citenamefont
  {Cederbaum}, \citenamefont {Zobeley},\ and\ \citenamefont
  {Tarantelli}}]{Cederbaum.1997}%
  \BibitemOpen
  \bibfield  {author} {\bibinfo {author} {\bibfnamefont {L.~S.}\ \bibnamefont
  {Cederbaum}}, \bibinfo {author} {\bibfnamefont {J.}~\bibnamefont {Zobeley}},\
  and\ \bibinfo {author} {\bibfnamefont {F.}~\bibnamefont {Tarantelli}},\
  }\href {https://doi.org/10.1103/PhysRevLett.79.4778} {\bibfield  {journal}
  {\bibinfo  {journal} {Physical Review Letters}\ }\textbf {\bibinfo {volume}
  {79}},\ \bibinfo {pages} {4778} (\bibinfo {year} {1997})}\BibitemShut
  {NoStop}%
\bibitem [{\citenamefont {Marburger}\ \emph {et~al.}(2003)\citenamefont
  {Marburger}, \citenamefont {Kugeler}, \citenamefont {Hergenhahn},\ and\
  \citenamefont {M{\"o}ller}}]{Marburger.2003}%
  \BibitemOpen
  \bibfield  {author} {\bibinfo {author} {\bibfnamefont {S.}~\bibnamefont
  {Marburger}}, \bibinfo {author} {\bibfnamefont {O.}~\bibnamefont {Kugeler}},
  \bibinfo {author} {\bibfnamefont {U.}~\bibnamefont {Hergenhahn}},\ and\
  \bibinfo {author} {\bibfnamefont {T.}~\bibnamefont {M{\"o}ller}},\ }\href
  {https://doi.org/10.1103/PhysRevLett.90.203401} {\bibfield  {journal}
  {\bibinfo  {journal} {Physical Review Letters}\ }\textbf {\bibinfo {volume}
  {90}},\ \bibinfo {pages} {203401} (\bibinfo {year} {2003})}\BibitemShut
  {NoStop}%
\bibitem [{\citenamefont {Jahnke}\ \emph {et~al.}(2004)\citenamefont {Jahnke},
  \citenamefont {Czasch}, \citenamefont {Sch{\"o}ffler}, \citenamefont
  {Sch{\"o}ssler}, \citenamefont {Knapp}, \citenamefont {K{\"a}sz},
  \citenamefont {Titze}, \citenamefont {Wimmer}, \citenamefont {Kreidi},
  \citenamefont {Grisenti}, \citenamefont {Staudte}, \citenamefont {Jagutzki},
  \citenamefont {Hergenhahn}, \citenamefont {Schmidt-B{\"o}cking},\ and\
  \citenamefont {D{\"o}rner}}]{Jahnke.2004}%
  \BibitemOpen
  \bibfield  {author} {\bibinfo {author} {\bibfnamefont {T.}~\bibnamefont
  {Jahnke}}, \bibinfo {author} {\bibfnamefont {A.}~\bibnamefont {Czasch}},
  \bibinfo {author} {\bibfnamefont {M.~S.}\ \bibnamefont {Sch{\"o}ffler}},
  \bibinfo {author} {\bibfnamefont {S.}~\bibnamefont {Sch{\"o}ssler}}, \bibinfo
  {author} {\bibfnamefont {A.}~\bibnamefont {Knapp}}, \bibinfo {author}
  {\bibfnamefont {M.}~\bibnamefont {K{\"a}sz}}, \bibinfo {author}
  {\bibfnamefont {J.}~\bibnamefont {Titze}}, \bibinfo {author} {\bibfnamefont
  {C.}~\bibnamefont {Wimmer}}, \bibinfo {author} {\bibfnamefont
  {K.}~\bibnamefont {Kreidi}}, \bibinfo {author} {\bibfnamefont {R.~E.}\
  \bibnamefont {Grisenti}}, \bibinfo {author} {\bibfnamefont {A.}~\bibnamefont
  {Staudte}}, \bibinfo {author} {\bibfnamefont {O.}~\bibnamefont {Jagutzki}},
  \bibinfo {author} {\bibfnamefont {U.}~\bibnamefont {Hergenhahn}}, \bibinfo
  {author} {\bibfnamefont {H.}~\bibnamefont {Schmidt-B{\"o}cking}},\ and\
  \bibinfo {author} {\bibfnamefont {R.}~\bibnamefont {D{\"o}rner}},\ }\href
  {https://doi.org/10.1103/PhysRevLett.93.163401} {\bibfield  {journal}
  {\bibinfo  {journal} {Physical Review Letters}\ }\textbf {\bibinfo {volume}
  {93}},\ \bibinfo {pages} {163401} (\bibinfo {year} {2004})}\BibitemShut
  {NoStop}%
\bibitem [{\citenamefont {Jahnke}\ \emph {et~al.}(2020)\citenamefont {Jahnke},
  \citenamefont {Hergenhahn}, \citenamefont {Winter}, \citenamefont
  {D{\"o}rner}, \citenamefont {Fr{\"u}hling}, \citenamefont {Demekhin},
  \citenamefont {Gokhberg}, \citenamefont {Cederbaum}, \citenamefont
  {Ehresmann}, \citenamefont {Knie},\ and\ \citenamefont
  {Dreuw}}]{Jahnke.2020}%
  \BibitemOpen
  \bibfield  {author} {\bibinfo {author} {\bibfnamefont {T.}~\bibnamefont
  {Jahnke}}, \bibinfo {author} {\bibfnamefont {U.}~\bibnamefont {Hergenhahn}},
  \bibinfo {author} {\bibfnamefont {B.}~\bibnamefont {Winter}}, \bibinfo
  {author} {\bibfnamefont {R.}~\bibnamefont {D{\"o}rner}}, \bibinfo {author}
  {\bibfnamefont {U.}~\bibnamefont {Fr{\"u}hling}}, \bibinfo {author}
  {\bibfnamefont {P.~V.}\ \bibnamefont {Demekhin}}, \bibinfo {author}
  {\bibfnamefont {K.}~\bibnamefont {Gokhberg}}, \bibinfo {author}
  {\bibfnamefont {L.~S.}\ \bibnamefont {Cederbaum}}, \bibinfo {author}
  {\bibfnamefont {A.}~\bibnamefont {Ehresmann}}, \bibinfo {author}
  {\bibfnamefont {A.}~\bibnamefont {Knie}},\ and\ \bibinfo {author}
  {\bibfnamefont {A.}~\bibnamefont {Dreuw}},\ }\href
  {https://doi.org/10.1021/acs.chemrev.0c00106} {\bibfield  {journal} {\bibinfo
   {journal} {Chemical Reviews}\ }\textbf {\bibinfo {volume} {120}},\ \bibinfo
  {pages} {11295} (\bibinfo {year} {2020})}\BibitemShut {NoStop}%
\bibitem [{\citenamefont {Mucke}\ \emph {et~al.}(2010)\citenamefont {Mucke},
  \citenamefont {Braune}, \citenamefont {Barth}, \citenamefont {F{\"o}rstel},
  \citenamefont {Lischke}, \citenamefont {Ulrich}, \citenamefont {Arion},
  \citenamefont {Becker}, \citenamefont {Bradshaw},\ and\ \citenamefont
  {Hergenhahn}}]{Mucke.2010}%
  \BibitemOpen
  \bibfield  {author} {\bibinfo {author} {\bibfnamefont {M.}~\bibnamefont
  {Mucke}}, \bibinfo {author} {\bibfnamefont {M.}~\bibnamefont {Braune}},
  \bibinfo {author} {\bibfnamefont {S.}~\bibnamefont {Barth}}, \bibinfo
  {author} {\bibfnamefont {M.}~\bibnamefont {F{\"o}rstel}}, \bibinfo {author}
  {\bibfnamefont {T.}~\bibnamefont {Lischke}}, \bibinfo {author} {\bibfnamefont
  {V.}~\bibnamefont {Ulrich}}, \bibinfo {author} {\bibfnamefont
  {T.}~\bibnamefont {Arion}}, \bibinfo {author} {\bibfnamefont
  {U.}~\bibnamefont {Becker}}, \bibinfo {author} {\bibfnamefont
  {A.}~\bibnamefont {Bradshaw}},\ and\ \bibinfo {author} {\bibfnamefont
  {U.}~\bibnamefont {Hergenhahn}},\ }\href {https://doi.org/10.1038/nphys1500}
  {\bibfield  {journal} {\bibinfo  {journal} {Nature Physics}\ }\textbf
  {\bibinfo {volume} {6}},\ \bibinfo {pages} {143} (\bibinfo {year}
  {2010})}\BibitemShut {NoStop}%
\bibitem [{\citenamefont {Hergenhahn}(2012)}]{Hergenhahn.2012}%
  \BibitemOpen
  \bibfield  {author} {\bibinfo {author} {\bibfnamefont {U.}~\bibnamefont
  {Hergenhahn}},\ }\href {https://doi.org/10.3109/09553002.2012.698031}
  {\bibfield  {journal} {\bibinfo  {journal} {International Journal of
  Radiation Biology}\ }\textbf {\bibinfo {volume} {88}},\ \bibinfo {pages}
  {871} (\bibinfo {year} {2012})}\BibitemShut {NoStop}%
\bibitem [{\citenamefont {Kimura}\ \emph
  {et~al.}(2013{\natexlab{a}})\citenamefont {Kimura}, \citenamefont {Fukuzawa},
  \citenamefont {Sakai}, \citenamefont {Mondal}, \citenamefont {Kukk},
  \citenamefont {Kono}, \citenamefont {Nagaoka}, \citenamefont {Tamenori},
  \citenamefont {Saito},\ and\ \citenamefont {Ueda}}]{Kimura.2013}%
  \BibitemOpen
  \bibfield  {author} {\bibinfo {author} {\bibfnamefont {M.}~\bibnamefont
  {Kimura}}, \bibinfo {author} {\bibfnamefont {H.}~\bibnamefont {Fukuzawa}},
  \bibinfo {author} {\bibfnamefont {K.}~\bibnamefont {Sakai}}, \bibinfo
  {author} {\bibfnamefont {S.}~\bibnamefont {Mondal}}, \bibinfo {author}
  {\bibfnamefont {E.}~\bibnamefont {Kukk}}, \bibinfo {author} {\bibfnamefont
  {Y.}~\bibnamefont {Kono}}, \bibinfo {author} {\bibfnamefont {S.}~\bibnamefont
  {Nagaoka}}, \bibinfo {author} {\bibfnamefont {Y.}~\bibnamefont {Tamenori}},
  \bibinfo {author} {\bibfnamefont {N.}~\bibnamefont {Saito}},\ and\ \bibinfo
  {author} {\bibfnamefont {K.}~\bibnamefont {Ueda}},\ }\href
  {https://doi.org/10.1103/PhysRevA.87.043414} {\bibfield  {journal} {\bibinfo
  {journal} {Physical Review A}\ }\textbf {\bibinfo {volume} {87}},\ \bibinfo
  {pages} {043414} (\bibinfo {year} {2013}{\natexlab{a}})}\BibitemShut
  {NoStop}%
\bibitem [{\citenamefont {Kimura}\ \emph
  {et~al.}(2013{\natexlab{b}})\citenamefont {Kimura}, \citenamefont {Fukuzawa},
  \citenamefont {Tachibana}, \citenamefont {Ito}, \citenamefont {Mondal},
  \citenamefont {Okunishi}, \citenamefont {Sch{\"o}ffler}, \citenamefont
  {Williams}, \citenamefont {Jiang}, \citenamefont {Tamenori}, \citenamefont
  {Saito},\ and\ \citenamefont {Ueda}}]{Kimura.2013b}%
  \BibitemOpen
  \bibfield  {author} {\bibinfo {author} {\bibfnamefont {M.}~\bibnamefont
  {Kimura}}, \bibinfo {author} {\bibfnamefont {H.}~\bibnamefont {Fukuzawa}},
  \bibinfo {author} {\bibfnamefont {T.}~\bibnamefont {Tachibana}}, \bibinfo
  {author} {\bibfnamefont {Y.}~\bibnamefont {Ito}}, \bibinfo {author}
  {\bibfnamefont {S.}~\bibnamefont {Mondal}}, \bibinfo {author} {\bibfnamefont
  {M.}~\bibnamefont {Okunishi}}, \bibinfo {author} {\bibfnamefont
  {M.}~\bibnamefont {Sch{\"o}ffler}}, \bibinfo {author} {\bibfnamefont
  {J.}~\bibnamefont {Williams}}, \bibinfo {author} {\bibfnamefont
  {Y.}~\bibnamefont {Jiang}}, \bibinfo {author} {\bibfnamefont
  {Y.}~\bibnamefont {Tamenori}}, \bibinfo {author} {\bibfnamefont
  {N.}~\bibnamefont {Saito}},\ and\ \bibinfo {author} {\bibfnamefont
  {K.}~\bibnamefont {Ueda}},\ }\href {https://doi.org/10.1021/jz4006674}
  {\bibfield  {journal} {\bibinfo  {journal} {The Journal of Physical Chemistry
  Letters}\ }\textbf {\bibinfo {volume} {4}},\ \bibinfo {pages} {1838}
  (\bibinfo {year} {2013}{\natexlab{b}})}\BibitemShut {NoStop}%
\bibitem [{\citenamefont {Miteva}\ \emph {et~al.}(2014)\citenamefont {Miteva},
  \citenamefont {Chiang}, \citenamefont {Koloren{\v{c}}}, \citenamefont
  {Kuleff}, \citenamefont {Cederbaum},\ and\ \citenamefont
  {Gokhberg}}]{Miteva.2014}%
  \BibitemOpen
  \bibfield  {author} {\bibinfo {author} {\bibfnamefont {T.}~\bibnamefont
  {Miteva}}, \bibinfo {author} {\bibfnamefont {Y.-C.}\ \bibnamefont {Chiang}},
  \bibinfo {author} {\bibfnamefont {P.}~\bibnamefont {Koloren{\v{c}}}},
  \bibinfo {author} {\bibfnamefont {A.~I.}\ \bibnamefont {Kuleff}}, \bibinfo
  {author} {\bibfnamefont {L.~S.}\ \bibnamefont {Cederbaum}},\ and\ \bibinfo
  {author} {\bibfnamefont {K.}~\bibnamefont {Gokhberg}},\ }\href
  {https://doi.org/10.1063/1.4898154} {\bibfield  {journal} {\bibinfo
  {journal} {The Journal of chemical physics}\ }\textbf {\bibinfo {volume}
  {141}},\ \bibinfo {pages} {164303} (\bibinfo {year} {2014})}\BibitemShut
  {NoStop}%
\bibitem [{\citenamefont {O'Keeffe}\ \emph {et~al.}(2013)\citenamefont
  {O'Keeffe}, \citenamefont {Ripani},  \citenamefont {Bolognesi},  \citenamefont 
{Coreno},  \citenamefont {Devetta},  \citenamefont {Callegari},  \citenamefont 
{Di Fraia} ,\citenamefont {Prince}, \citenamefont {Richter},  \citenamefont 
{Alagia},  \citenamefont {Kivim{\"{a}}ki},\ and\
  \citenamefont {Avaldi}}]{OKeeffe2014}%
  \BibitemOpen
  \bibfield  {author} {\bibinfo {author} {\bibfnamefont {P.}~\bibnamefont
  {O'Keeffe}}, \bibinfo {author} {\bibfnamefont {E.}~\bibnamefont
  {Ripani}}, \bibinfo {author} {\bibfnamefont {P.}~\bibnamefont
  {Bolognesi}}, \bibinfo {author} {\bibfnamefont {M.}~\bibnamefont
  {Coreno}}, \bibinfo {author} {\bibfnamefont {M.}~\bibnamefont
  {Devetta}}, \bibinfo {author} {\bibfnamefont {C.}~\bibnamefont
  {Callegari}}, \bibinfo {author} {\bibfnamefont {M.}~\bibnamefont
  {Di Fraia}}, \bibinfo {author} {\bibfnamefont {K.~C.}~\bibnamefont
  {Prince}}, \bibinfo {author} {\bibfnamefont {R.}~\bibnamefont
  {Richter}}, \bibinfo {author} {\bibfnamefont {M.}~\bibnamefont
  {Alagia}}, \bibinfo {author} {\bibfnamefont {A.}\ \bibnamefont
  {Kivim{\"{a}}ki}},\ and\ \bibinfo {author} {\bibfnamefont {L.}\ \bibnamefont
  {Avaldi}},\ }\href {https://doi.org/ 10.1088/1742-6596/488/2/022015} {\bibfield
  {journal} {\bibinfo  {journal} {J. Phys. Chem. Lett.}\ }\textbf {\bibinfo {volume} {4}},\
  \bibinfo {pages} {1797} (\bibinfo {year} {2013})}\BibitemShut {NoStop}%
\bibitem [{\citenamefont {Malerz}\ \emph {et~al.}(2021)\citenamefont {Malerz},
  \citenamefont {Trinter}, \citenamefont {Hergenhahn}, \citenamefont {Ghrist},
  \citenamefont {Ali}, \citenamefont {Nicolas}, \citenamefont {Saak},
  \citenamefont {Richter}, \citenamefont {Hartweg}, \citenamefont {Nahon},
  \citenamefont {Lee}, \citenamefont {Goy}, \citenamefont {Neumark},
  \citenamefont {Meijer}, \citenamefont {Wilkinson}, \citenamefont {Winter},\
  and\ \citenamefont {Th{\"u}rmer}}]{Malerz.2021}%
  \BibitemOpen
  \bibfield  {author} {\bibinfo {author} {\bibfnamefont {S.}~\bibnamefont
  {Malerz}}, \bibinfo {author} {\bibfnamefont {F.}~\bibnamefont {Trinter}},
  \bibinfo {author} {\bibfnamefont {U.}~\bibnamefont {Hergenhahn}}, \bibinfo
  {author} {\bibfnamefont {A.}~\bibnamefont {Ghrist}}, \bibinfo {author}
  {\bibfnamefont {H.}~\bibnamefont {Ali}}, \bibinfo {author} {\bibfnamefont
  {C.}~\bibnamefont {Nicolas}}, \bibinfo {author} {\bibfnamefont {C.-M.}\
  \bibnamefont {Saak}}, \bibinfo {author} {\bibfnamefont {C.}~\bibnamefont
  {Richter}}, \bibinfo {author} {\bibfnamefont {S.}~\bibnamefont {Hartweg}},
  \bibinfo {author} {\bibfnamefont {L.}~\bibnamefont {Nahon}}, \bibinfo
  {author} {\bibfnamefont {C.}~\bibnamefont {Lee}}, \bibinfo {author}
  {\bibfnamefont {C.}~\bibnamefont {Goy}}, \bibinfo {author} {\bibfnamefont
  {D.~M.}\ \bibnamefont {Neumark}}, \bibinfo {author} {\bibfnamefont
  {G.}~\bibnamefont {Meijer}}, \bibinfo {author} {\bibfnamefont
  {I.}~\bibnamefont {Wilkinson}}, \bibinfo {author} {\bibfnamefont
  {B.}~\bibnamefont {Winter}},\ and\ \bibinfo {author} {\bibfnamefont
  {S.}~\bibnamefont {Th{\"u}rmer}},\ }\href
  {https://doi.org/10.1039/D1CP00430A} {\bibfield  {journal} {\bibinfo
  {journal} {Physical Chemistry Chemical Physics}\ }\textbf {\bibinfo {volume}
  {23}},\ \bibinfo {pages} {8246} (\bibinfo {year} {2021})}\BibitemShut
  {NoStop}%
\bibitem [{\citenamefont {Abid}\ \emph {et~al.}(2021)\citenamefont {Abid},
  \citenamefont {Mailhiot}, \citenamefont {Boudjemia}, \citenamefont
  {Pelimanni}, \citenamefont {Milosavljevi{\'c}}, \citenamefont {Saak},
  \citenamefont {Huttula}, \citenamefont {Bj{\"o}rneholm},\ and\ \citenamefont
  {Patanen}}]{Abid.2021}%
  \BibitemOpen
  \bibfield  {author} {\bibinfo {author} {\bibfnamefont {A.~R.}\ \bibnamefont
  {Abid}}, \bibinfo {author} {\bibfnamefont {M.}~\bibnamefont {Mailhiot}},
  \bibinfo {author} {\bibfnamefont {N.}~\bibnamefont {Boudjemia}}, \bibinfo
  {author} {\bibfnamefont {E.}~\bibnamefont {Pelimanni}}, \bibinfo {author}
  {\bibfnamefont {A.~R.}\ \bibnamefont {Milosavljevi{\'c}}}, \bibinfo {author}
  {\bibfnamefont {C.-M.}\ \bibnamefont {Saak}}, \bibinfo {author}
  {\bibfnamefont {M.}~\bibnamefont {Huttula}}, \bibinfo {author} {\bibfnamefont
  {O.}~\bibnamefont {Bj{\"o}rneholm}},\ and\ \bibinfo {author} {\bibfnamefont
  {M.}~\bibnamefont {Patanen}},\ }\href {https://doi.org/10.1039/d0ra08943e}
  {\bibfield  {journal} {\bibinfo  {journal} {RSC Advances}\ }\textbf {\bibinfo
  {volume} {11}},\ \bibinfo {pages} {2103} (\bibinfo {year}
  {2021})}\BibitemShut {NoStop}%
\bibitem [{\citenamefont {Yang}\ \emph {et~al.}(2020)\citenamefont {Yang},
  \citenamefont {Liu}, \citenamefont {Feng}, \citenamefont {Qian},
  \citenamefont {Kao}, \citenamefont {Ha}, \citenamefont {Hahn}, \citenamefont
  {Seguin}, \citenamefont {Tsige}, \citenamefont {Yang}, \citenamefont
  {Zavadil}, \citenamefont {Persson},\ and\ \citenamefont {Guo}}]{Yang.2020}%
  \BibitemOpen
  \bibfield  {author} {\bibinfo {author} {\bibfnamefont {F.}~\bibnamefont
  {Yang}}, \bibinfo {author} {\bibfnamefont {Y.-S.}\ \bibnamefont {Liu}},
  \bibinfo {author} {\bibfnamefont {X.}~\bibnamefont {Feng}}, \bibinfo {author}
  {\bibfnamefont {K.}~\bibnamefont {Qian}}, \bibinfo {author} {\bibfnamefont
  {L.~C.}\ \bibnamefont {Kao}}, \bibinfo {author} {\bibfnamefont
  {Y.}~\bibnamefont {Ha}}, \bibinfo {author} {\bibfnamefont {N.~T.}\
  \bibnamefont {Hahn}}, \bibinfo {author} {\bibfnamefont {T.~J.}\ \bibnamefont
  {Seguin}}, \bibinfo {author} {\bibfnamefont {M.}~\bibnamefont {Tsige}},
  \bibinfo {author} {\bibfnamefont {W.}~\bibnamefont {Yang}}, \bibinfo {author}
  {\bibfnamefont {K.~R.}\ \bibnamefont {Zavadil}}, \bibinfo {author}
  {\bibfnamefont {K.~A.}\ \bibnamefont {Persson}},\ and\ \bibinfo {author}
  {\bibfnamefont {J.}~\bibnamefont {Guo}},\ }\href
  {https://doi.org/10.1039/D0RA05905F} {\bibfield  {journal} {\bibinfo
  {journal} {RSC Advances}\ }\textbf {\bibinfo {volume} {10}},\ \bibinfo
  {pages} {27315} (\bibinfo {year} {2020})}\BibitemShut {NoStop}%
\bibitem [{\citenamefont {Rubensson}\ \emph {et~al.}(1994)\citenamefont
  {Rubensson}, \citenamefont {Eisebitt}, \citenamefont {Nicodemus},
  \citenamefont {B{\"o}ske},\ and\ \citenamefont {Eberhardt}}]{Rubensson.1994}%
  \BibitemOpen
  \bibfield  {author} {\bibinfo {author} {\bibfnamefont {J.-E.}\ \bibnamefont
  {Rubensson}}, \bibinfo {author} {\bibfnamefont {S.}~\bibnamefont {Eisebitt}},
  \bibinfo {author} {\bibfnamefont {M.}~\bibnamefont {Nicodemus}}, \bibinfo
  {author} {\bibfnamefont {T.}~\bibnamefont {B{\"o}ske}},\ and\ \bibinfo
  {author} {\bibfnamefont {W.}~\bibnamefont {Eberhardt}},\ }\href
  {https://doi.org/10.1103/PhysRevB.50.9035} {\bibfield  {journal} {\bibinfo
  {journal} {Physical Review B}\ }\textbf {\bibinfo {volume} {50}},\ \bibinfo
  {pages} {9035} (\bibinfo {year} {1994})}\BibitemShut {NoStop}%
\bibitem [{\citenamefont {Pokapanich}\ \emph {et~al.}(2011)\citenamefont
  {Pokapanich}, \citenamefont {Kryzhevoi}, \citenamefont {Ottosson},
  \citenamefont {Svensson}, \citenamefont {Cederbaum}, \citenamefont
  {{\"O}hrwall},\ and\ \citenamefont {Bj{\"o}rneholm}}]{Pokapanich.2011}%
  \BibitemOpen
  \bibfield  {author} {\bibinfo {author} {\bibfnamefont {W.}~\bibnamefont
  {Pokapanich}}, \bibinfo {author} {\bibfnamefont {N.~V.}\ \bibnamefont
  {Kryzhevoi}}, \bibinfo {author} {\bibfnamefont {N.}~\bibnamefont {Ottosson}},
  \bibinfo {author} {\bibfnamefont {S.}~\bibnamefont {Svensson}}, \bibinfo
  {author} {\bibfnamefont {L.~S.}\ \bibnamefont {Cederbaum}}, \bibinfo {author}
  {\bibfnamefont {G.}~\bibnamefont {{\"O}hrwall}},\ and\ \bibinfo {author}
  {\bibfnamefont {O.}~\bibnamefont {Bj{\"o}rneholm}},\ }\href
  {https://doi.org/10.1021/ja203430s} {\bibfield  {journal} {\bibinfo
  {journal} {Journal of the American Chemical Society}\ }\textbf {\bibinfo
  {volume} {133}},\ \bibinfo {pages} {13430} (\bibinfo {year}
  {2011})}\BibitemShut {NoStop}%
\bibitem [{\citenamefont {Ottosson}\ \emph {et~al.}(2012)\citenamefont
  {Ottosson}, \citenamefont {{\"O}hrwall},\ and\ \citenamefont
  {Bj{\"o}rneholm}}]{Ottosson.2012}%
  \BibitemOpen
  \bibfield  {author} {\bibinfo {author} {\bibfnamefont {N.}~\bibnamefont
  {Ottosson}}, \bibinfo {author} {\bibfnamefont {G.}~\bibnamefont
  {{\"O}hrwall}},\ and\ \bibinfo {author} {\bibfnamefont {O.}~\bibnamefont
  {Bj{\"o}rneholm}},\ }\href {https://doi.org/10.1016/j.cplett.2012.05.051}
  {\bibfield  {journal} {\bibinfo  {journal} {Chemical Physics Letters}\
  }\textbf {\bibinfo {volume} {543}},\ \bibinfo {pages} {1} (\bibinfo {year}
  {2012})}\BibitemShut {NoStop}%
\bibitem [{\citenamefont {Dupuy}\ \emph {et~al.}(2024)\citenamefont
  {Dupuy}, \citenamefont {Buttersack}, \citenamefont {Trinter}, \citenamefont {Richter}, 
\citenamefont {Gholami}, \citenamefont {Bj{\"o}rneholm}, \citenamefont {Hergenhahn}, 
\citenamefont {Winter},\ and\ \citenamefont {Bluhm}}]{Dupuy.2024}%
\BibitemOpen
\bibfield  {author} {\bibinfo {author} {\bibfnamefont {R.}~\bibnamefont
  {Dupuy}}, \bibinfo {author} {\bibfnamefont {T.}~\bibnamefont
{Buttersack}}, \bibinfo {author} {\bibfnamefont {F.}~\bibnamefont
{Trinter}}, \bibinfo {author} {\bibfnamefont {C.}~\bibnamefont
{Richter}}, \bibinfo {author} {\bibfnamefont {S.}~\bibnamefont
{Gholami}}, \bibinfo {author} {\bibfnamefont {O.}~\bibnamefont
{Bj{\"o}rneholm }}, \bibinfo {author} {\bibfnamefont {U.}~\bibnamefont
{Hergenhahn}}, \bibinfo {author} {\bibfnamefont {B.}~\bibnamefont
{Winter}},\ and\ \bibinfo {author} {\bibfnamefont {H.}~\bibnamefont
  {Bluhm}},\ }\href { Article https://doi.org/10.1038/s41467-024-51417-3} {\bibfield  
{journal} {\bibinfo  {journal} {Nature Communications}\ 
 }\textbf {\bibinfo {volume} {15}},\ \bibinfo {pages} {6923} (\bibinfo {year}  
{2024})}\BibitemShut {NoStop}%
\bibitem [{\citenamefont {Winter}\ \emph {et~al.}(2023)\citenamefont {Winter},
  \citenamefont {Th{\"u}rmer},\ and\ \citenamefont {Wilkinson}}]{Winter.2023}%
  \BibitemOpen
  \bibfield  {author} {\bibinfo {author} {\bibfnamefont {B.}~\bibnamefont
  {Winter}}, \bibinfo {author} {\bibfnamefont {S.}~\bibnamefont
  {Th{\"u}rmer}},\ and\ \bibinfo {author} {\bibfnamefont {I.}~\bibnamefont
  {Wilkinson}},\ }\href {https://doi.org/10.1021/acs.accounts.2c00548}
  {\bibfield  {journal} {\bibinfo  {journal} {Accounts of Chemical Research}\
  }\textbf {\bibinfo {volume} {56}},\ \bibinfo {pages} {77} (\bibinfo {year}
  {2023})}\BibitemShut {NoStop}%
\bibitem [{\citenamefont {Jalilehvand}\ \emph {et~al.}(2001)\citenamefont
  {Jalilehvand}, \citenamefont {Sp{\aa}ngberg}, \citenamefont {Lindqvist-Reis},
  \citenamefont {Hermansson}, \citenamefont {Persson},\ and\ \citenamefont
  {Sandstr{\"o}m}}]{Jalilehvand.2001}%
  \BibitemOpen
  \bibfield  {author} {\bibinfo {author} {\bibfnamefont {F.}~\bibnamefont
  {Jalilehvand}}, \bibinfo {author} {\bibfnamefont {D.}~\bibnamefont
  {Sp{\aa}ngberg}}, \bibinfo {author} {\bibfnamefont {P.}~\bibnamefont
  {Lindqvist-Reis}}, \bibinfo {author} {\bibfnamefont {K.}~\bibnamefont
  {Hermansson}}, \bibinfo {author} {\bibfnamefont {I.}~\bibnamefont
  {Persson}},\ and\ \bibinfo {author} {\bibfnamefont {M.}~\bibnamefont
  {Sandstr{\"o}m}},\ }\href {https://doi.org/10.1021/ja001533a} {\bibfield
  {journal} {\bibinfo  {journal} {Journal of the American Chemical Society}\
  }\textbf {\bibinfo {volume} {123}},\ \bibinfo {pages} {431} (\bibinfo {year}
  {2001})}\BibitemShut {NoStop}%
\bibitem [{SM()}]{SM}%
  \BibitemOpen
  \href@noop {} {\bibinfo  {journal} {Supplemental Material}\ }\BibitemShut
  {NoStop}%
\bibitem [{\citenamefont {Blo{\ss}}\ \emph {et~al.}(2024)\citenamefont
  {Blo{\ss}}, \citenamefont {Trinter},  \citenamefont {Unger},  \citenamefont 
  {Zindel},  \citenamefont {Honisch},  \citenamefont {Viehmann},  \citenamefont 
  {Kiefer} ,\citenamefont {Prince}, \citenamefont {Marder},  \citenamefont 
  {K{\"{u}}stner-Wetekam},  \citenamefont {Heikura}, \citenamefont 
  {Cederbaum}, \citenamefont {Bj{\"{o}}rneholm}, \citenamefont 
  {Hergenhahn}, \citenamefont {Ehresmann}, \ and\  \citenamefont
  {Hans}}]{Blo.2024}%
  \BibitemOpen
  \bibfield  {author} {\bibinfo {author} {\bibfnamefont {D.}~\bibnamefont
  {Blo{\ss}}}, \bibinfo {author} {\bibfnamefont {F.}~\bibnamefont
  {Trinter}}, \bibinfo {author} {\bibfnamefont {I.}~\bibnamefont
  {Unger}}, \bibinfo {author} {\bibfnamefont {C.}~\bibnamefont
  {Zindel}}, \bibinfo {author} {\bibfnamefont {C.}~\bibnamefont
  {Honisch}}, \bibinfo {author} {\bibfnamefont {J.}~\bibnamefont
  {Viehmann}}, \bibinfo {author} {\bibfnamefont {N.}~\bibnamefont
  {Kiefer}}, \bibinfo {author} {\bibfnamefont {L.}~\bibnamefont
  {Marder}}, \bibinfo {author} {\bibfnamefont {C.}~\bibnamefont
  {K{\"{u}}stner-Wetekam}}, \bibinfo {author} {\bibfnamefont {E.}~\bibnamefont
  {Heikura}},\bibinfo {author} {\bibfnamefont {L.~S.}~\bibnamefont
  {Cederbaum}},\bibinfo {author} {\bibfnamefont {O.}~\bibnamefont
  {Bj{\"{o}}rneholm}}, \bibinfo {author} {\bibfnamefont {U.}~\bibnamefont
  {Hergenhahn}}, \bibinfo {author} {\bibfnamefont {A.}\ \bibnamefont
  {Ehresmann}},\ and\ \bibinfo {author} {\bibfnamefont {A.}\ \bibnamefont
  {Hans}},\ }\href { https://doi.org/10.1038/s41467-024-48687-2} {\bibfield
  {journal} {\bibinfo  {journal} { Nature Communications}\ }\textbf {\bibinfo {volume} {15}},\
  \bibinfo {pages} {4594} (\bibinfo {year} {2024})}\BibitemShut {NoStop}%
\bibitem [{\citenamefont {Gopakumar}\ \emph {et~al.}(2023)\citenamefont
  {Gopakumar}, \citenamefont {Unger}, \citenamefont {Slav{\'i}{\v{c}}ek},
  \citenamefont {Hergenhahn}, \citenamefont {{\"O}hrwall}, \citenamefont
  {Malerz}, \citenamefont {C{\'e}olin}, \citenamefont {Trinter}, \citenamefont
  {Winter}, \citenamefont {Wilkinson}, \citenamefont {Caleman}, \citenamefont
  {Muchov{\'a}},\ and\ \citenamefont {Bj{\"o}rneholm}}]{Gopakumar.2023}%
  \BibitemOpen
  \bibfield  {author} {\bibinfo {author} {\bibfnamefont {G.}~\bibnamefont
  {Gopakumar}}, \bibinfo {author} {\bibfnamefont {I.}~\bibnamefont {Unger}},
  \bibinfo {author} {\bibfnamefont {P.}~\bibnamefont {Slav{\'i}{\v{c}}ek}},
  \bibinfo {author} {\bibfnamefont {U.}~\bibnamefont {Hergenhahn}}, \bibinfo
  {author} {\bibfnamefont {G.}~\bibnamefont {{\"O}hrwall}}, \bibinfo {author}
  {\bibfnamefont {S.}~\bibnamefont {Malerz}}, \bibinfo {author} {\bibfnamefont
  {D.}~\bibnamefont {C{\'e}olin}}, \bibinfo {author} {\bibfnamefont
  {F.}~\bibnamefont {Trinter}}, \bibinfo {author} {\bibfnamefont
  {B.}~\bibnamefont {Winter}}, \bibinfo {author} {\bibfnamefont
  {I.}~\bibnamefont {Wilkinson}}, \bibinfo {author} {\bibfnamefont
  {C.}~\bibnamefont {Caleman}}, \bibinfo {author} {\bibfnamefont
  {E.}~\bibnamefont {Muchov{\'a}}},\ and\ \bibinfo {author} {\bibfnamefont
  {O.}~\bibnamefont {Bj{\"o}rneholm}},\ }\href
  {https://doi.org/10.1038/s41557-023-01302-1} {\bibfield  {journal} {\bibinfo
  {journal} {Nature Chemistry}\ }\textbf {\bibinfo {volume} {15}},\ \bibinfo
  {pages} {1408} (\bibinfo {year} {2023})}\BibitemShut {NoStop}%
\bibitem [{\citenamefont {Stumpf}\ \emph {et~al.}(2016)\citenamefont {Stumpf},
  \citenamefont {Gokhberg},\ and\ \citenamefont {Cederbaum}}]{Stumpf.2016}%
  \BibitemOpen
  \bibfield  {author} {\bibinfo {author} {\bibfnamefont {V.}~\bibnamefont
  {Stumpf}}, \bibinfo {author} {\bibfnamefont {K.}~\bibnamefont {Gokhberg}},\
  and\ \bibinfo {author} {\bibfnamefont {L.~S.}\ \bibnamefont {Cederbaum}},\
  }\href {https://doi.org/10.1038/NCHEM.2429} {\bibfield  {journal} {\bibinfo
  {journal} {Nature Chemistry}\ }\textbf {\bibinfo {volume} {8}},\ \bibinfo
  {pages} {237} (\bibinfo {year} {2016})}\BibitemShut {NoStop}%
\bibitem [{\citenamefont {Orlando}\ \emph {et~al.}(2008)\citenamefont
  {Orlando}, \citenamefont {Oh}, \citenamefont {Chen},\ and\ \citenamefont
  {Aleksandrov}}]{Orlando.2008}%
  \BibitemOpen
  \bibfield  {author} {\bibinfo {author} {\bibfnamefont {T.~M.}\ \bibnamefont
  {Orlando}}, \bibinfo {author} {\bibfnamefont {D.}~\bibnamefont {Oh}},
  \bibinfo {author} {\bibfnamefont {Y.}~\bibnamefont {Chen}},\ and\ \bibinfo
  {author} {\bibfnamefont {A.~B.}\ \bibnamefont {Aleksandrov}},\ }\href
  {https://doi.org/10.1063/1.2907722} {\bibfield  {journal} {\bibinfo
  {journal} {The Journal of chemical physics}\ }\textbf {\bibinfo {volume}
  {128}},\ \bibinfo {pages} {195102} (\bibinfo {year} {2008})}\BibitemShut
  {NoStop}%
\bibitem [{\citenamefont {Winter}\ and\ \citenamefont
  {Faubel}(2006)}]{Winter.2006}%
  \BibitemOpen
  \bibfield  {author} {\bibinfo {author} {\bibfnamefont {B.}~\bibnamefont
  {Winter}}\ and\ \bibinfo {author} {\bibfnamefont {M.}~\bibnamefont
  {Faubel}},\ }\href {https://doi.org/10.1021/cr040381p} {\bibfield  {journal}
  {\bibinfo  {journal} {Chemical Reviews}\ }\textbf {\bibinfo {volume} {106}},\
  \bibinfo {pages} {1176} (\bibinfo {year} {2006})}\BibitemShut {NoStop}%
\bibitem [{\citenamefont {Malerz}\ \emph {et~al.}(2022)\citenamefont {Malerz},
  \citenamefont {Haak}, \citenamefont {Trinter}, \citenamefont {Stephansen},
  \citenamefont {Kolbeck}, \citenamefont {Pohl}, \citenamefont {Hergenhahn},
  \citenamefont {Meijer},\ and\ \citenamefont {Winter}}]{Malerz.2022}%
  \BibitemOpen
  \bibfield  {author} {\bibinfo {author} {\bibfnamefont {S.}~\bibnamefont
  {Malerz}}, \bibinfo {author} {\bibfnamefont {H.}~\bibnamefont {Haak}},
  \bibinfo {author} {\bibfnamefont {F.}~\bibnamefont {Trinter}}, \bibinfo
  {author} {\bibfnamefont {A.~B.}\ \bibnamefont {Stephansen}}, \bibinfo
  {author} {\bibfnamefont {C.}~\bibnamefont {Kolbeck}}, \bibinfo {author}
  {\bibfnamefont {M.}~\bibnamefont {Pohl}}, \bibinfo {author} {\bibfnamefont
  {U.}~\bibnamefont {Hergenhahn}}, \bibinfo {author} {\bibfnamefont
  {G.}~\bibnamefont {Meijer}},\ and\ \bibinfo {author} {\bibfnamefont
  {B.}~\bibnamefont {Winter}},\ }\href {https://doi.org/10.1063/5.0072346}
  {\bibfield  {journal} {\bibinfo  {journal} {The Review of Scientific
  Instruments}\ }\textbf {\bibinfo {volume} {93}},\ \bibinfo {pages} {015101}
  (\bibinfo {year} {2022})}\BibitemShut {NoStop}%
\bibitem [{\citenamefont {Viefhaus}\ \emph {et~al.}(2013)\citenamefont
  {Viefhaus}, \citenamefont {Scholz}, \citenamefont {Deinert}, \citenamefont
  {Glaser}, \citenamefont {Ilchen}, \citenamefont {Seltmann}, \citenamefont
  {Walter},\ and\ \citenamefont {Siewert}}]{Viefhaus.2013}%
  \BibitemOpen
  \bibfield  {author} {\bibinfo {author} {\bibfnamefont {J.}~\bibnamefont
  {Viefhaus}}, \bibinfo {author} {\bibfnamefont {F.}~\bibnamefont {Scholz}},
  \bibinfo {author} {\bibfnamefont {S.}~\bibnamefont {Deinert}}, \bibinfo
  {author} {\bibfnamefont {L.}~\bibnamefont {Glaser}}, \bibinfo {author}
  {\bibfnamefont {M.}~\bibnamefont {Ilchen}}, \bibinfo {author} {\bibfnamefont
  {J.}~\bibnamefont {Seltmann}}, \bibinfo {author} {\bibfnamefont
  {P.}~\bibnamefont {Walter}},\ and\ \bibinfo {author} {\bibfnamefont
  {F.}~\bibnamefont {Siewert}},\ }\href
  {https://doi.org/10.1016/j.nima.2012.10.110} {\bibfield  {journal} {\bibinfo
  {journal} {Nuclear Instruments and Methods in Physics Research Section A}\
  }\textbf {\bibinfo {volume} {710}},\ \bibinfo {pages} {151} (\bibinfo {year}
  {2013})}\BibitemShut {NoStop}%
\bibitem [{\citenamefont {Pohl}\ \emph {et~al.}(2017)\citenamefont {Pohl},
  \citenamefont {Richter}, \citenamefont {Lugovoy}, \citenamefont {Seidel},
  \citenamefont {Slav{\'i}{\v{c}}ek}, \citenamefont {Aziz}, \citenamefont
  {Abel}, \citenamefont {Winter},\ and\ \citenamefont
  {Hergenhahn}}]{Pohl.2017}%
  \BibitemOpen
  \bibfield  {author} {\bibinfo {author} {\bibfnamefont {M.~N.}\ \bibnamefont
  {Pohl}}, \bibinfo {author} {\bibfnamefont {C.}~\bibnamefont {Richter}},
  \bibinfo {author} {\bibfnamefont {E.}~\bibnamefont {Lugovoy}}, \bibinfo
  {author} {\bibfnamefont {R.}~\bibnamefont {Seidel}}, \bibinfo {author}
  {\bibfnamefont {P.}~\bibnamefont {Slav{\'i}{\v{c}}ek}}, \bibinfo {author}
  {\bibfnamefont {E.~F.}\ \bibnamefont {Aziz}}, \bibinfo {author}
  {\bibfnamefont {B.}~\bibnamefont {Abel}}, \bibinfo {author} {\bibfnamefont
  {B.}~\bibnamefont {Winter}},\ and\ \bibinfo {author} {\bibfnamefont
  {U.}~\bibnamefont {Hergenhahn}},\ }\href
  {https://doi.org/10.1021/acs.jpcb.7b06061} {\bibfield  {journal} {\bibinfo
  {journal} {The Journal of Physical Chemistry B}\ }\textbf {\bibinfo {volume}
  {121}},\ \bibinfo {pages} {7709} (\bibinfo {year} {2017})}\BibitemShut
  {NoStop}%
\bibitem [{\citenamefont {Preobrajenski}\ \emph {et~al.}(2023)\citenamefont
  {Preobrajenski}, \citenamefont {Generalov}, \citenamefont {{\"O}hrwall},
  \citenamefont {Tchaplyguine}, \citenamefont {Tarawneh}, \citenamefont
  {Appelfeller}, \citenamefont {Frampton},\ and\ \citenamefont
  {Walsh}}]{Preobrajenski.2023}%
  \BibitemOpen
  \bibfield  {author} {\bibinfo {author} {\bibfnamefont {A.}~\bibnamefont
  {Preobrajenski}}, \bibinfo {author} {\bibfnamefont {A.}~\bibnamefont
  {Generalov}}, \bibinfo {author} {\bibfnamefont {G.}~\bibnamefont
  {{\"O}hrwall}}, \bibinfo {author} {\bibfnamefont {M.}~\bibnamefont
  {Tchaplyguine}}, \bibinfo {author} {\bibfnamefont {H.}~\bibnamefont
  {Tarawneh}}, \bibinfo {author} {\bibfnamefont {S.}~\bibnamefont
  {Appelfeller}}, \bibinfo {author} {\bibfnamefont {E.}~\bibnamefont
  {Frampton}},\ and\ \bibinfo {author} {\bibfnamefont {N.}~\bibnamefont
  {Walsh}},\ }\href {https://doi.org/10.1107/S1600577523003429} {\bibfield
  {journal} {\bibinfo  {journal} {Journal of Synchrotron Radiation}\ }\textbf
  {\bibinfo {volume} {30}},\ \bibinfo {pages} {831} (\bibinfo {year}
  {2023})}\BibitemShut {NoStop}%
\end{thebibliography}%
\bibliographystyle{apsrev4-2}

\begin{acknowledgments}
\section{Acknowledgements}
We acknowledge DESY (Hamburg, Germany), a member of the Helmholtz Association HGF, for the provision of experimental facilities and allocation of beamtime for proposal I-20211422. We are grateful for the excellent support from the PETRA~III P04 beamline staff. The authors acknowledge MAX IV Laboratory for time on the beamline FlexPES under Proposal 20230898. Research conducted at MAX IV, a Swedish national user facility, is supported by the Swedish Research Council under contract 2018-07152, the Swedish Governmental Agency for Innovation Systems under contract 2018-04969, and Formas under contract 2019-02496. This work was supported by the German Federal Ministry of Education and Research (BMBF) through projects 05K22RK2 – GPhaseCC and 05K22RK1 – TRANSALP and SFB 1319 ELCH, funded by the Deutsche Forschungsgemeinschaft (DFG; project No. 328961117). We also acknowledge the scientific exchange and support of the Centre for Molecular Water Science (CMWS). F.\,T. acknowledges funding by the Deutsche Forschungsgemeinschaft (DFG, German Research Foundation) - Project 509471550, Emmy Noether Programme and acknowledges support by the MaxWater initiative of the Max-Planck-Gesellschaft. L.\,S.\,C. gratefully acknowledges financial support by the European Research Council (ERC) (Advanced Investigator Grant No. 692657). O.\,B. acknowledges support the Swedish Research Council VR through project 2023-04346.
\end{acknowledgments}

\end{document}


\title{Supplemental Material for manuscript\\
``Site- and Energy-Selective Low-Energy Electron Emission by X-Rays in Aqueous Phase''}

\author{Dana Blo\ss}
\email{dana-bloss@uni-kassel.de}
	\affiliation{Institut f\"ur Physik und CINSaT, Universit\"at Kassel, Heinrich-Plett-Straße\,40, 34132 Kassel, Germany}

\author{R\'{e}mi Dupuy}
	\affiliation{Laboratoire de Chimie Physique - Mati\`{e}re et Rayonnement, Sorbonne Universit\'{e}, CNRS, LCP-MR, 75005 Paris Cedex 05, France}
	
\author{Florian Trinter}
	\affiliation{Fritz-Haber-Institut der Max-Planck-Gesellschaft, Faradayweg 4-6, 14195 Berlin, Germany}
	\affiliation{Institut f\"ur Kernphysik, Goethe-Universit\"at Frankfurt, Max-von-Laue-Straße\,1, 60438 Frankfurt am Main, Germany}
	
\author{Isaak Unger}
	\affiliation{Department of Physics and Astronomy, Uppsala University, Box 516, 75120 Uppsala, Sweden}
	
\author{Noelle Walsh}
	\affiliation{MAX IV Laboratory, Lund University, Box 118, 22100 Lund, Sweden}
	
\author{\\Gunnar \"{O}hrwall}
	\affiliation{MAX IV Laboratory, Lund University, Box 118, 22100 Lund, Sweden}

\author{Niklas Golchert}
	\affiliation{Institut f\"ur Physik und CINSaT, Universit\"at Kassel, Heinrich-Plett-Straße\,40, 34132 Kassel, Germany}

\author{Gabriel Klassen}
	\affiliation{Institut f\"ur Physik und CINSaT, Universit\"at Kassel, Heinrich-Plett-Straße\,40, 34132 Kassel, Germany}
\author{Adrian Krone}
	\affiliation{Institut f\"ur Physik und CINSaT, Universit\"at Kassel, Heinrich-Plett-Straße\,40, 34132 Kassel, Germany}

\author{Yusaku Terao}
		\affiliation{Institut f\"ur Physik und CINSaT, Universit\"at Kassel, Heinrich-Plett-Straße\,40, 34132 Kassel, Germany}
		
\author{\\Johannes H. Viehmann}
	\affiliation{Institut f\"ur Physik und CINSaT, Universit\"at Kassel, Heinrich-Plett-Straße\,40, 34132 Kassel, Germany}
	
\author{Lasse W\"{u}lfing}
	\affiliation{Fakult\"{a}t Physik, Technische Universit\"{a}t Dortmund, Maria-Goeppert-Mayer-Straße 2, 44227 Dortmund, Germany}

\author{Clemens Richter}
	\affiliation{Fritz-Haber-Institut der Max-Planck-Gesellschaft, Faradayweg 4-6, 14195 Berlin, Germany}

\author{Tillmann Buttersack}
	\affiliation{Fritz-Haber-Institut der Max-Planck-Gesellschaft, Faradayweg 4-6, 14195 Berlin, Germany}

\author{Lorenz S. Cederbaum}
	\affiliation{Theoretische Chemie, Institut f\"ur Physikalische Chemie, Universit\"at Heidelberg, Im Neuenheimer Feld 229, 69120 Heidelberg, Germany}
	
\author{\\Uwe Hergenhahn}
	\affiliation{Fritz-Haber-Institut der Max-Planck-Gesellschaft, Faradayweg 4-6, 14195 Berlin, Germany}

\author{Olle Bj\"orneholm}
	\affiliation{Department of Physics and Astronomy, Uppsala University, Box 516, 75120 Uppsala, Sweden}
	
\author{Arno Ehresmann}
	\affiliation{Institut f\"ur Physik und CINSaT, Universit\"at Kassel, Heinrich-Plett-Straße\,40, 34132 Kassel, Germany}

\author{Andreas Hans}
\email{hans@physik.uni-kassel.de}
	\affiliation{Institut f\"ur Physik und CINSaT, Universit\"at Kassel, Heinrich-Plett-Straße\,40, 34132 Kassel, Germany}

\date{\today}
\maketitle

\section{Identification of the resonant Auger electrons in the time-of-flight spectra}
In the coincidence experiment using the magnetic-bottle time-of-flight spectrometer, the absolute energy resolution decreases for faster electrons, i.e., for electrons of higher kinetic energy. It is therefore essential to identify the spectator resonant Auger electrons unambiguously, which are subsequently used as a filtering condition in the coincidence data set. For that purpose, we measured electron spectra on the Ca$^{2+}$ $2p\rightarrow 3d$ resonances and above the ionization threshold and compared them to the high-resolution spectra obtained with the hemispherical analyzer. An exemplary spectrum is shown in Fig.~\ref{fig:spectraonoff}. A clear peak can be identified at 280-290~eV kinetic energy in the spectrum on resonance, which is less pronounced and slightly shifted to lower energies in the spectrum above threshold. Comparison to the spectra in Fig.~3 in the main manuscript clearly allows the assignment of this feature to spectator resonant Auger electrons. The spectra in Fig.~\ref{fig:spectraonoff} comprise events in which only one electron has been detected.

\renewcommand{\thefigure}{S1}
\begin{figure}[h!]
\includegraphics[width=.46\textwidth]{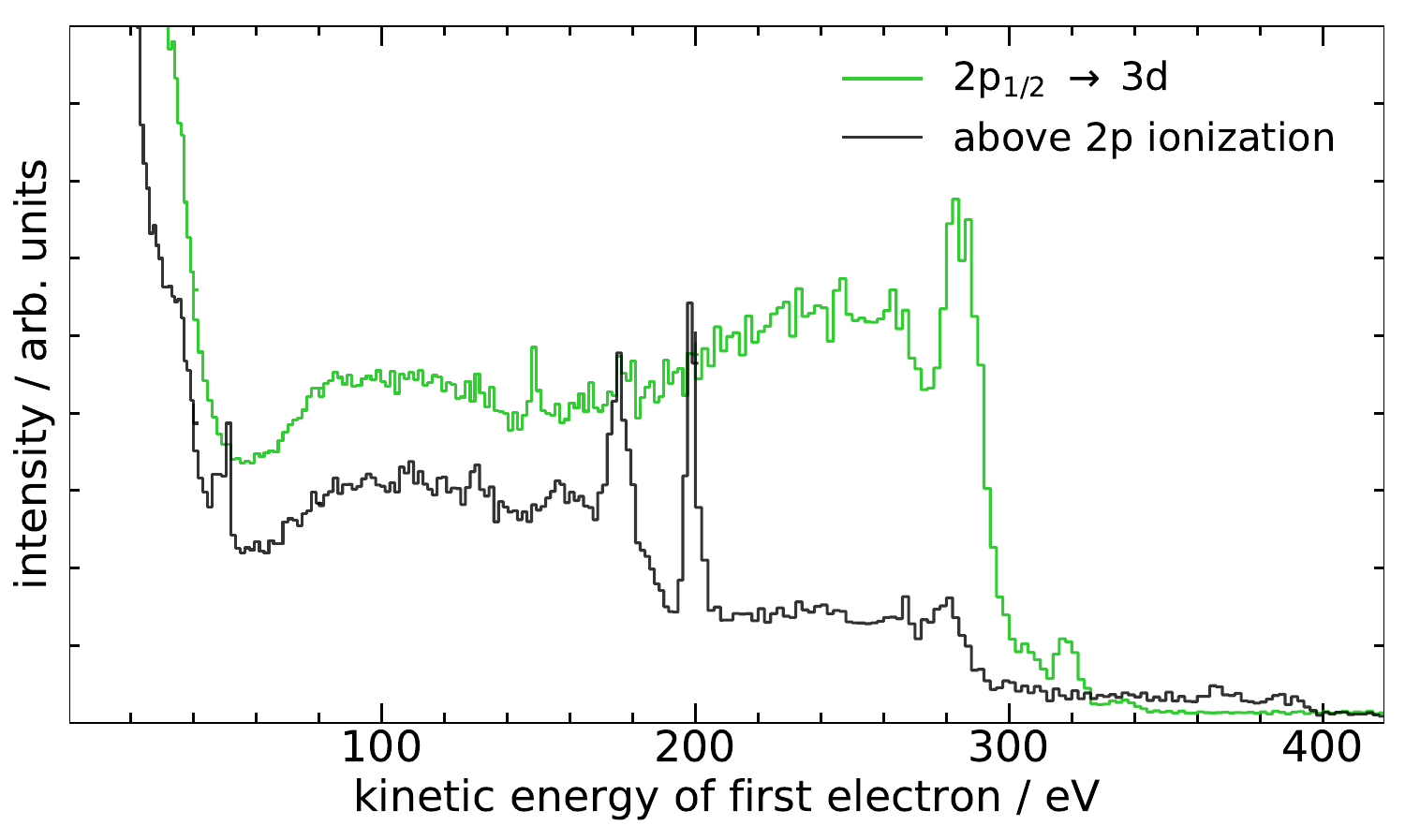}
  \caption{Electron spectra of a 4~M CaCl$_2$ solution recorded on the $2p_{\nicefrac{1}{2}}\rightarrow 3d$ resonance (green trace) and above the $2p$ ionization threshold (405~eV, black trace). In the spectrum on resonance, the spectator resonant Auger electrons can readily be identified at 280-290~eV kinetic energy. In addition, the participator Auger emission can be observed at about 320~eV, see also Fig.~3 of the main manuscript. Features in the 150-200~eV kinetic-energy range can be assigned to Cl$^-$ $2p$ photoelectrons and Auger electrons, and the peak at about 50~eV in the above-threshold spectrum to Ca$^{2+}$ $2p$ photoelectrons. The intensity of the spectra has been scaled for better presentability, a quantitative comparison is not meaningful due to the different absorption cross sections.}
  \label{fig:spectraonoff}
\end{figure}

The two-dimensional electron-electron coincidence maps recorded on the $2p_{\nicefrac{3}{2}}\rightarrow 3d$ and $2p_{\nicefrac{1}{2}}\rightarrow 3d$ resonances are displayed in Fig.~\ref{fig:eemaps}. Such maps are histograms showing all double-electron coincidence events plotted with the kinetic energy of the first electron on the $x$ axis and the kinetic energy of the second electron on the $y$ axis. The low-energy electron (LEE) tail characteristic for all experiments on liquids is very prominent. All electrons in coincidence with a spectator resonant Auger electron appear in a column at 280-290~eV on the $x$~axis. The feature assigned to ICD following resonant Auger decay (RA-ICD) is indicated by arrows in both maps.

\renewcommand{\thefigure}{S2}
\begin{figure}[h!]
\includegraphics[width=.75\textwidth]{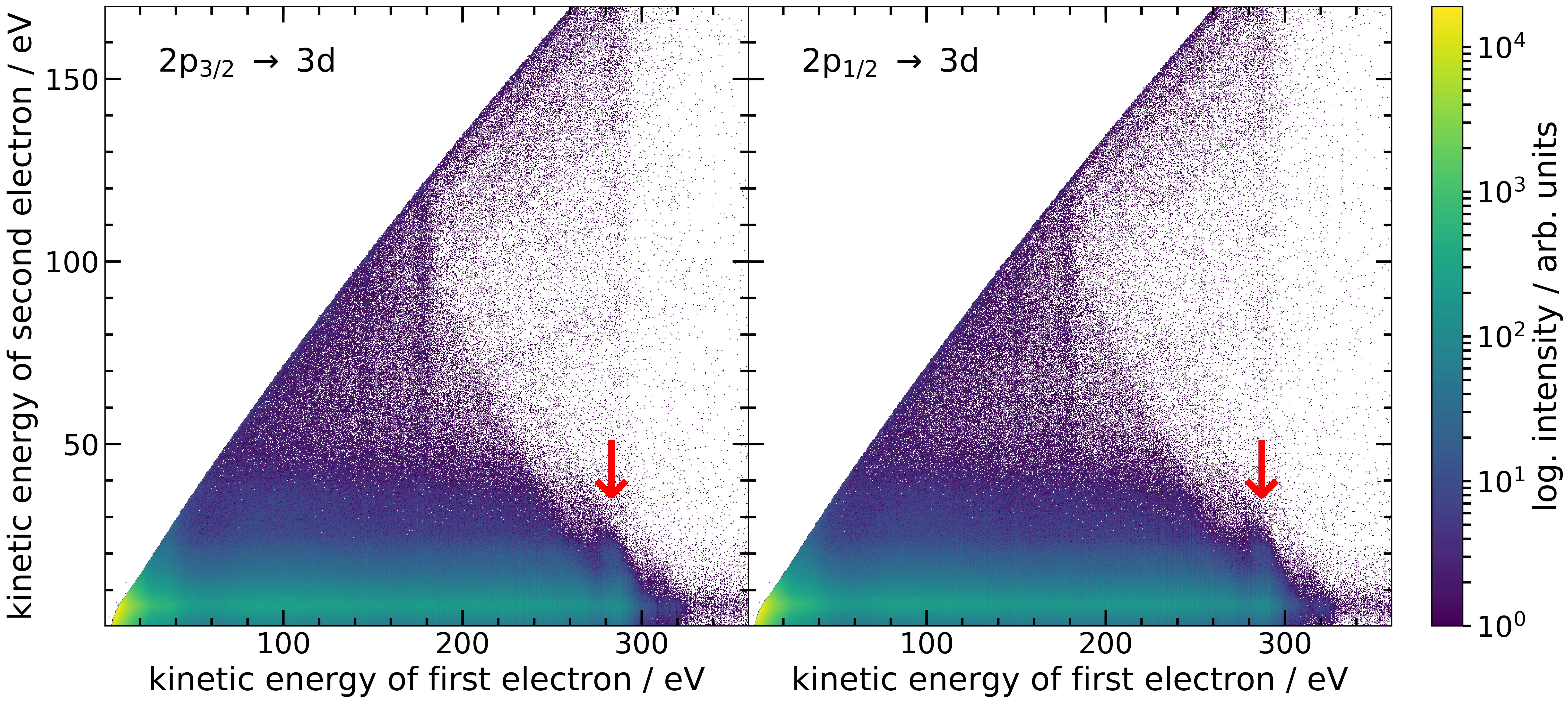}
  \caption{Electron-electron coincidence maps recorded on the $2p_{\nicefrac{3}{2}}\rightarrow 3d$ (left) and $2p_{\nicefrac{1}{2}}\rightarrow 3d$ (right) resonances. The red arrows indicate the feature attributed to RA-ICD for each case.}
  \label{fig:eemaps}
\end{figure}

\section{Estimate of the relevance of random coincidences}
\renewcommand{\thefigure}{S3}
\begin{figure}[h!]
\includegraphics[width=.55\textwidth]{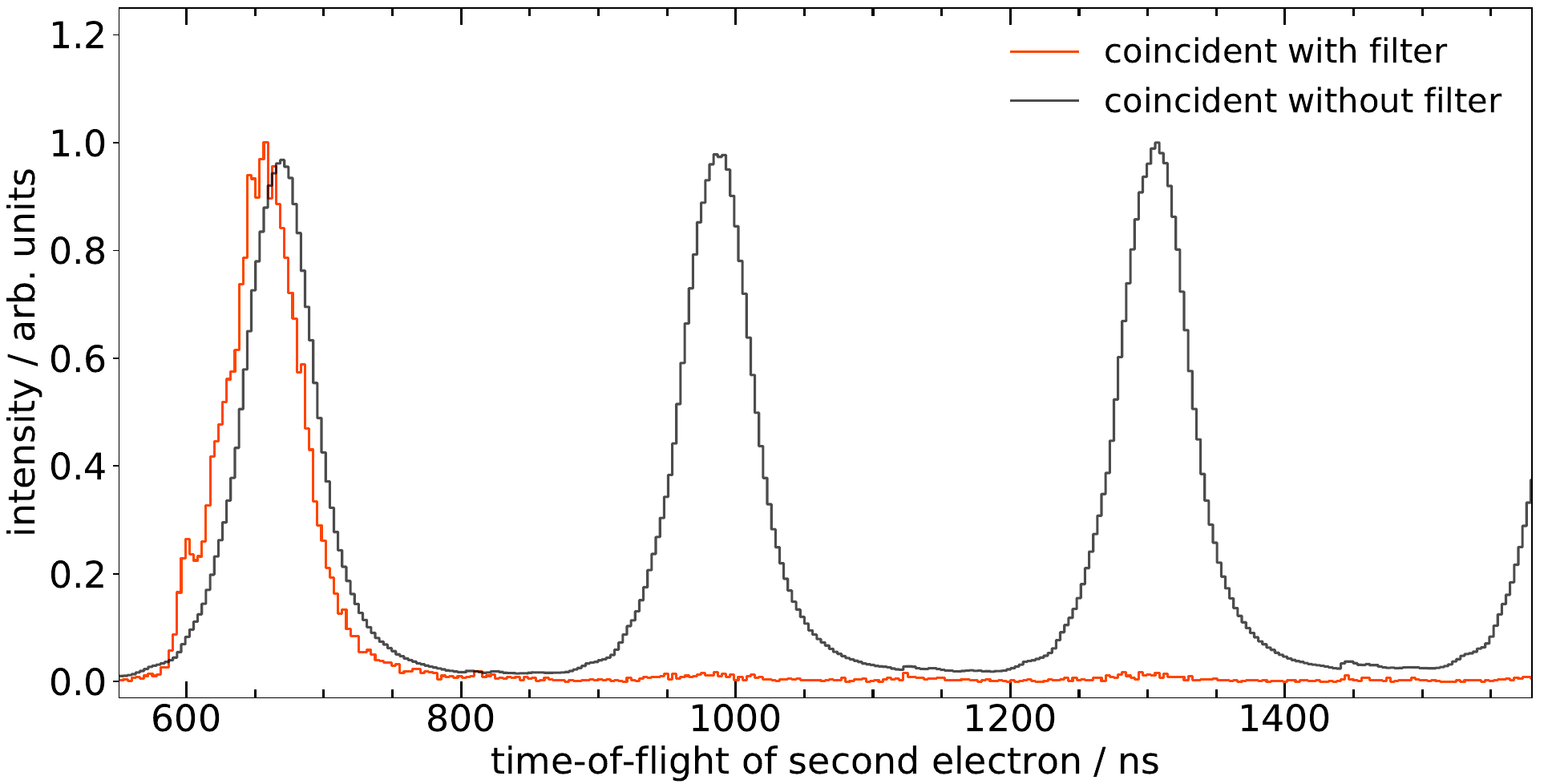}
  \caption{Time-of-flight spectra of the second detected electron in two-electron-coincidence events on the $2p_{\nicefrac{1}{2}}\rightarrow 3d$ resonance. The black trace shows all second electrons as measured, yielding an expected periodicity of 320~ns. For the red trace, a filter condition was applied for the first electron detected being the resonant Auger electron. The periodicity disappears, indicating that random coincidences are negligible.}
  \label{fig:random}
\end{figure}

If more than one interaction event occurs within a single exciting-photon pulse, this can cause random coincidences. This can be avoided by keeping the event rate $f_{event}$ much lower than the repetition rate $f_{exc}$ of the excitation, i.e., $f_{event}\ll f_{exc}$. For double-electron coincidences the occurrence of random coincidences can be monitored by setting the acquisition time of the time-to-digital converter (TDC) longer than at least double the temporal spacing between two consecutive exciting-photon pulses. If then a filter condition is applied to the first of two detected electrons, random coincidences will appear as a periodic pattern in the time-of-flight spectrum of the second detected electron with the same periodicity as the excitation pulses (320~ns in the present case). During the experiment at the FlexPES beamline, the acquisition time was set to 1600~ns, i.e., covering 5~pulses. In Fig.~\ref{fig:random}, we show the time-of-flight spectrum of the second electron detected in two-electron coincidence events without any further filter and with the condition that the first detected electron is the Ca$^{2+}$ resonant Auger electron (for the $2p_{\nicefrac{1}{2}}\rightarrow 3d$ resonance). As expected, without any filter the spectrum shows periodic structures repeating every 320~ns. Under the filter condition, all but one feature disappear.  The remaining feature originates from true and random coincidences with the electrons within the filter window. Random coincidences would appear as a periodic structure, which is practically absent. From this we conclude that random coincidences can be neglected in the data treatment.

\section{Correlation between the kinetic energies of the resonant Auger electrons and the ICD electrons}

A weak, diagonally shaped feature due to RA-ICD feature can be observed in Fig.~\ref{fig:eemaps}. It can straightforwardly be interpreted as a constant sum of resonant Auger electron and ICD electron. The slower the Auger electron, the more internal energy is still stored in the ion, and the more energy can be released in the transition to the ground state. Consequently, the maximum energy of the ICD electron increases. We emphasize this by dividing the coincidence filter condition, which has been used to obtain the spectra in Fig.~4 of the main manuscript, in two parts, the low- and high-energy parts of the spectator resonant Auger peak. Both corresponding RA-ICD spectra are shown in Fig.~\ref{fig:ICDvsAuger} for the $2p_{\nicefrac{3}{2}}\rightarrow 3d$ resonance at 349.25~eV. As expected, a weak but significant shift of intensity in the RA-ICD feature according to the selected range of the Auger peak can be observed.

\renewcommand{\thefigure}{S4}
\begin{figure}[h!]
\includegraphics[width=.48\textwidth]{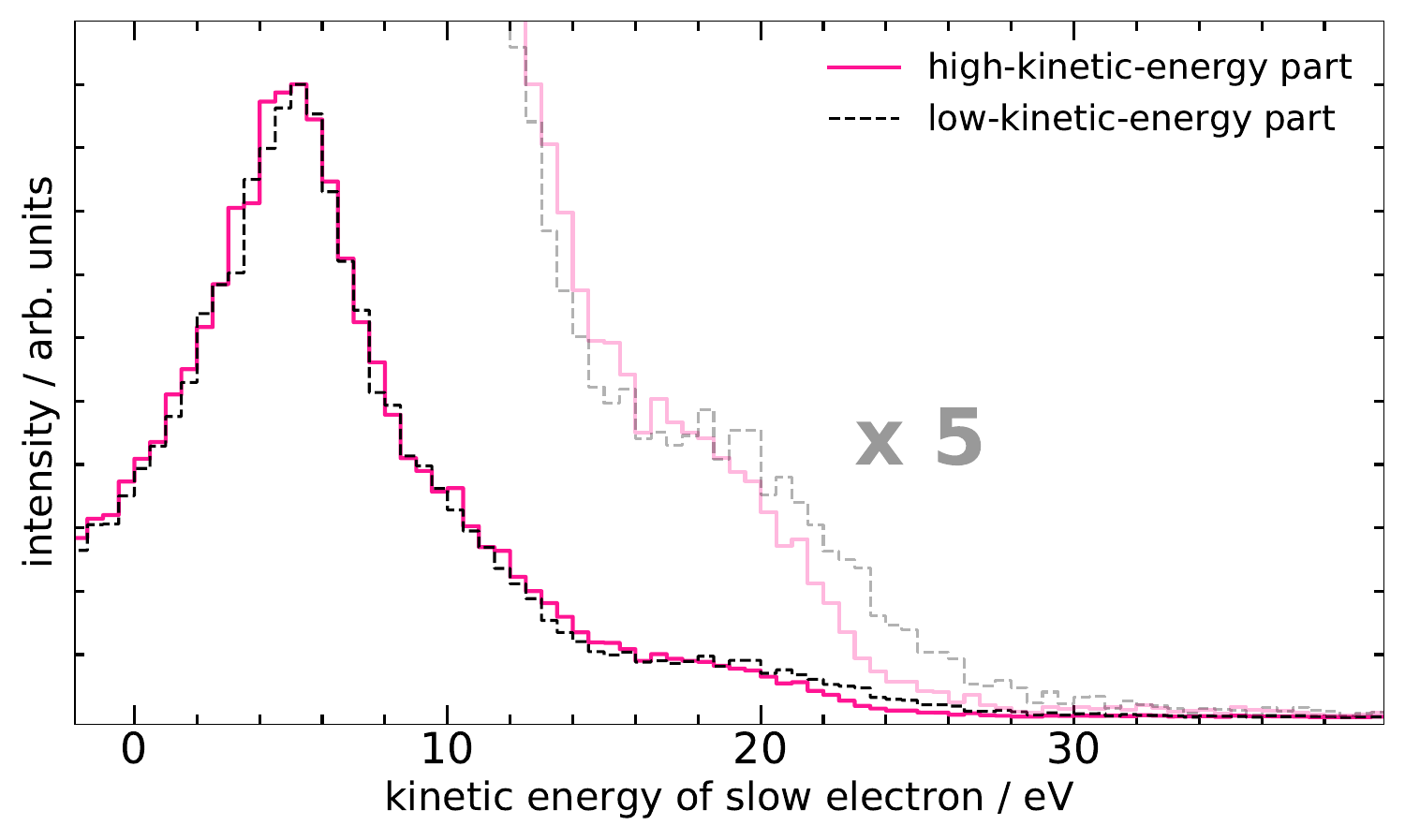}
  \caption{RA-ICD electron spectra on the $2p_{\nicefrac{3}{2}}\rightarrow 3d$ resonance at 349.25~eV for narrow coincidence conditions containing the low- or high-kinetic-energy part of the resonant Auger spectrum. A correlation between the kinetic energy of the Auger electrons and the RA-ICD electrons can be observed, namely, the lower the Auger electron kinetic energy, the higher the RA-ICD electron energy and \textit{vice versa}. This correlation describes the release of a constant amount of energy in the two-step transition from the inner-shell vacancy to the final state. }
  \label{fig:ICDvsAuger}
\end{figure}

\bibliography{references}
\bibliographystyle{apsrev4-2}